\def\kms{\ifmmode{\hbox{km~s}^{-1}}\else{km~s$^{-1}$}\fi}
\documentstyle[12pt,aaspp4]{article}
\begin{document}
\title {\bf ACTION PRINCIPLE SOLUTIONS FOR GALAXY MOTIONS WITHIN
3000 KM S$^{-1}$}

\author{Edward J. Shaya}
\affil {Department of Physics, University of Maryland, College Park,
Maryland 20742}

\author {P.J.E. Peebles}
\affil{ Physics Department, Princeton University, Princeton, NJ 08544}

\centerline{\&}

\author {R. Brent Tully}
\affil {Institute for Astronomy, University of Hawaii, Honolulu, HI
96822}

\begin{abstract}

The numerical action variational principle is used to find fully
nonlinear solutions for the orbits of the mass tracers given their
present redshifts and angular positions and the cosmological boundary
condition that the peculiar velocities are small at high redshift. A
solution predicts the distances of the mass tracers, and is tested by a
comparison with measured distances. The current numerical results use
289 luminosity-linewidth distance measurements designed to be close to
unbiased.  A catalog of 1138 tracers approximates the luminosity
distribution of galaxies in the vicinity of the Local Supercluster, at
redshifts $cz < 3000$ \kms. These mass tracers include groups with
crossing times less than the Hubble time and isolated galaxies. In this
preliminary computation, we assign each mass tracer the same
mass-to-light ratio $M/L$. The tracer masses are fixed by apparent
magnitudes and model distances. The only two free parameters in this
model are $M/L$ and the expansion time $t_0$. The measure of merit of a
solution is the sum of the mean square differences between the predicted
and observed distance moduli. In the 3000~\kms\ sample, this reduced
$\chi^2$ statistic has a well-defined minimum value at $M/L=175$ and
$t_0 = 10.0$ Gyr, and $\chi^2$ at the minimum is about 1.29 times the
value expected from just the standard deviation of the distance
measurements. We have tested for the effect of the mass at greater
distance by using the positions of Abell clusters as a model for the
large-scale mass distribution.  This external mass model reduces the
minimum value of $\chi^2$ by about 10\% ($\sim 1 \sigma$). The value of
the cosmological density parameter $\Omega_0$ is determined by the
global mean mass-to-light ratio. Our preliminary analysis yields
$\Omega_0= 0.17 \pm 0.10$ at one standard deviation.  A tighter bound is
expected to come out of a larger sample of measured distances now
available.

\end{abstract}


\section{Introduction}

If we could reconstruct the orbits that galaxies followed to arrive at
their present observed positions and velocities, there is much we would
learn about the distribution of mass and the initial conditions of
fluctuations.  A discussion of how to achieve this reconstruction is
begun here.  This analysis of the deviations from homogeneous Hubble
flow involves four steps: (1) establish a catalog of positions and
redshifts of objects that represent the present-day mass distribution;
(2) solve for orbits of these mass tracers that result in their present
angular positions and redshifts; (3) measure distances to a sample of
the mass tracers, and (4) use a statistical measure of the differences
between the observed distances and the distances in the orbit solution
to optimize the free parameters.  The program combines previous
discussions of the galaxy distances and redshifts (Shaya, Tully, \&
Pierce 1992: STP92; Tully, Shaya, \& Pierce 1992: TSP92) and a numerical
method of computation of the orbits (Peebles, 1989, 1990, 1994a: P89,
P90, P94). Our purpose for this paper is to present the considerations
that lead to our choices of the astronomical and numerical techniques,
and to show some first numerical results from the analysis of the
observations.

Numerical solutions to the N-body problem in step (2) require six
independent constraints on the orbit of each mass tracer. In a
Robertson-Walker line element, three of the constraints can be set by
the assumption that the peculiar velocities were negligible compared to
the rates of expansion between nearest neighbors at earliest times.  Two
additional constraints can be set by the present location of tracers in
the plane of the sky, and the last constraint can be the observed
redshift.  These three observables can be measured to high accuracy.
Theoretically, then, if one had a complete and accurate catalog of mass
tracers out to a reasonable distance, with positions in celestial
coordinates, redshifts, and masses, and one knew the age of the
universe, one would know the discrete set of possible galaxy orbits
since formation. In the process, one would predict the present distances
and transverse velocities. The action variational principle of classical
mechanics is readily adapted to the computation of orbits given
locations in the sky plus distances.  This method has been applied to
the Local Group of galaxies and its immediate neighbors (P89, P90, P94;
Dunn \& Laflamme 1993). In \S 2.1 we describe a modification that
determines the orbits and predicts the present distances given the
present redshifts.

We do not have the perfect catalog just described, of course, and there
are two main obstacles to creating it. First, the zone of avoidance
causes optical catalogs to be incomplete. We describe in \S 3.1, the use
one can make of observations in the infrared and radio to provide
information on the galaxy distribution at low galactic latitudes. Our
conclusion is that the zone of avoidance is a diminishing problem.
Second, non-luminous mass need not cluster with the galaxies. If the
dark mass has low pressure, it is very likely that the mass clusters
with the galaxies on large scales, because the gravitational instability
of the expanding universe draws galaxies and dark matter together in the
same way. If so, the issue would be the scale of significant segregation
of galaxies and dark mass. We argue in \S 6.2 that if galaxies and mass
were significantly segregated on scales of a few megaparsecs we would
expect to detect the effect in the relative motions of the galaxies in
and near the Local Group.  A mass component with a broader coherence
length might be present, and might be detected by a study of the
relative motions of the more widely separated mass tracers. There also
is the very real possibility that galaxies in denser regions have larger
mass-to-light ratios, meaning there is a mass component that is more
tightly clustered than the galaxies. Clearly, much remains to be learned
from modeling the relation between starlight and the underlying mass.

Step (3) in the program requires a distance measure that is unbiased or
has biases one understands and can remove. In \S 4 it is explained why
we think our biases are minimal. Differences of observed and predicted
distance moduli provide a measure of the fit of the model to the
observations, and we apply the luminosity-linewidth (Tully-Fisher)
method in a way that is expected to yield unbiased estimates of
individual distance moduli with a standard deviation $\delta\mu\sim
0.4$, independent of distance. This uncertainty $\delta\mu$ is in
reasonable agreement with the residuals found in the numerical fit in \S
5, an indication that we have a statistical measure of goodness-of-fit
that behaves in a well-understood way.

The main result of the computation is the measure of a $\chi^2$ fit as
a function of the independent parameters.  In the model we focus on
here, the cosmological constant vanishes and all mass tracers have the
same mass-to-light ratio $M/L$, so the independent parameters are $M/L$
and the expansion time $t_0$.  There are two independent parameters
rather than the single parameter that appears in some studies because
we do not assume the mean density of our sample agrees with the cosmic
mean (or, what is equivalent, that the mean rate of expansion of the
sample agrees with the background Hubble flow).  Rather, as a
secondary product that does not govern the fit, we derive the mean
density of the sample and find that it is close to the cosmic mean
when the parameters are close to the minimum of $\chi^2$.  The
relevant parameter from the cosmological model is $t_0$ because, in
the absence of a cosmological constant or other smoothly distributed
component in the stress-energy tensor, time is the only locally
observable cosmological parameter.  If this simplicity seems
unexpected recall that in general relativity theory local physics is
unaffected by the expansion of the universe except insofar as the expansion
sets the time $t_0$.

To summarize this important point: rather than work in the domain of
mass perturbations on a mean density and peculiar velocities upon a
mean Hubble expansion, we consider total masses that act over a
specified time to generate global observed velocities.


\section{The Action Variational Principle Methodology}

The numerical action procedure in P89, P90, and P94 yields orbits with
given present positions for the mass tracers treated as point particles.
The position of each tracer as a function of time is supposed to trace
the motion of the center of mass of the components out of which it is
assembled, and it is assumed that the gravitational effect of these
components on the motion of other mass elements at high redshift is well
approximated by replacing the distributed mass with the single particle
at the center of mass.  Some tests of this picture are discussed in \S 6.2.

The mass tracers have comoving positions $x_i(t)$. The equation
of motion, assuming a matter-dominated universe with negligible
nongravitational forces throughout the relevant
time span, can be derived from the paths at extrema and saddle points of
the action
\begin{equation}
S = \int^{t_0}_0 \Bigl[\sum{m_ia^2 \over 2} \Bigl({dx_i \over dt}\Bigr)^2
 + {G \over a}  \sum_{i \ne j} {m_im_j \over |x_i - x_j|}
 + {2 \over 3} \pi G \bar \rho a^2\sum m_ix_i^2\Bigr]dt,
\end{equation}
where $\bar \rho$ is the background density.

We plan to consider two families of cosmological models, those with zero
cosmological constant $\Lambda$ and those with a non-zero $\Lambda$ set
so that the universe has zero space curvature. The latter is consistent
with most interpretations of inflation as an explanation of the
large-scale homogeneity of the observable universe. A model with nonzero
values of both $\Lambda$ and space curvature seems more unlikely because
it would require a double coincidence of the epochs of life on Earth,
the transition of the expansion rate to curvature-dominance, and the
transition to $\Lambda$-dominance.

In the two families of models, time intervals are
\begin{equation}
H_0dt = {a^{1/2}da \over [\Omega_0 + a^p(1-\Omega_0)]^{1/2}},
\label{eq_worldmod}
\end{equation}
where p=1 if $\Lambda=0$ and p=3 if the model is cosmologically flat.
In most of the numerical results presented in this paper we assume
$\Lambda =0$, but we mention some trials that indicate the
cosmologically flat model would yield quite similar results with
larger expansion times.

The orbits are approximated by parametric functions:
\begin{equation}
x_i^\alpha(t) = x_i^\alpha(t_0) + \sum_n C^\alpha_{i,n}f_n(t),
\end{equation}
where $\alpha$ denotes the spatial component and the $f_n$ are a
set of functions satisfying the boundary conditions. The choice is not
critical; we use
\begin{equation}
f_n = (1-a)^{s-n}a^n\Bigl[ {s! \over n!(s-n)!}\Bigr],\quad   0 \le n <
s.
\label{eq_fn}
\end{equation}
The solutions to the equation of motion occur
where the gradients with respect to the coefficients of the trial
function vanish,
\begin{equation} {\delta S \over \delta C^\alpha_{i,n} }  = 0.
\end{equation}

In general, there may be many solutions consistent with the boundary
conditions. We choose the mass tracers by grouping galaxies into systems
that are not likely to have crossed so, apart from the classical
triple-valued regions, the interesting solution is the minimum of the
action. We find this minimum by walking down the gradient of the action
until the sum of the squares of the gradients is below a target three or
four orders of magnitude below the value in an initial trial that had
all coefficients set to zero and the distances set to the initial
guesses discussed below.

\subsection{Iterations to Observed Redshifts}

The only three well-determined phase space coordinates in large catalogs
are the two coordinates of projected positions and the radial
velocities;  distances are at best only approximately known. We know of
two ways to adapt the algorithm to predict distances given redshifts. A
variational method (Peebles 1994a) requires a matrix inversion that
seems impractical for a large sample. The alternative used here finds
present distances given the redshifts by the following iterative
procedure. The first step is to make initial guesses for the distances
of the tracers and solve for the orbits. The orbit solutions in general
are not acceptably close to the given velocities. Therefore, the local
gradients in the model velocities with respect to the given present
distances are determined numerically.  All of the distances along the
lines joining mass tracers and  the Local Group are adjusted from the
initial guesses by positive 0.1 Mpc.  After this small distance
change, new solutions are calculated.  For each mass tracer, the small
change in model redshift is compared to the difference between model and
observed redshift, and an estimate is made of the distance shift that
will bring the model in line with the observed redshift.  The actual
distance adjustment is the smaller of half the estimated shift or 1 Mpc.
Limiting the distance steps in this way prevents overshoot to
qualitatively new configurations in any one step, which is important to
prevent jumps to new branches of multiple-value regions.

The procedure is repeated with increasing accuracy by decreasing the
target value of the sum of the squares of the gradients. After about 10
iterations there is usually excellent agreement between calculated and
observed redshifts for all but one or two mass tracers.  The exceptions
usually are tracers that have been caught at a triple-value region and
need to get beyond it.  An increase in the minimum step size for one
iteration near the end of the computation usually lets these tracers
escape and then resettle at a better location.

\subsection{Numerical Details}

A softening parameter in the inverse square force law is set to
$10^{-2}$ times the sample radius or $c = 30\hbox{ km s}^{-1} = 0.3
h^{-1}$ Mpc. The force softening is useful in the initial iterations,
but after the mass tracers have nearly settled to their final paths $c$
may be reduced to zero with no significant consequences since there are
no close interactions in this sample, by design (see \S 3.1).

To check the orbit solutions from the action principle we use a
conventional N-body code to integrate the motions forward in time from
initial positions and velocities taken from the earliest time step in
the action variational solution.  We use 90 time steps in this
integration. Good agreement between the N-body and action orbits are
normally found for 5 parameters ($s=4$ in equation~[\ref{eq_fn}]). As to be
expected, when there are still small errors in the early time motion,
perhaps because the target value for the sum of the squares of the
gradients has not been set low enough, the present positions from the
N-body code may be quite different from the true values. Yet, the action
principle solution usually is reasonably accurate. This test therefore
requires one to calculate to greater precision in the action method than
one really cares, but it is worth the effort to have an accurate N-body
confirmation.

We ran additional tests with the mass-to-light ratio set to $M/L = 1$
because that case can be checked analytically. Because $M/L$ is so small
this case is equivalent to free Hubble expansion in a region that is
empty save for test particles of negligible mass. One trial used a
background cosmological model with $\Omega_0 = 0.6$ and $\Lambda = 0$,
and in another trial $\Lambda$ was set to make the universe
cosmologically flat.  All 1138 particles were used in these tests.  The
velocities of particles were within $\sim 1\%$ of theoretical
expectations.

For 1138 mass tracers, there are $1138 \times 5 \times 3 = 17,070$
coefficients to vary to minimize the action. The integrals are
calculated with 30 time steps, resulting in approximately 3\% accuracy
for peculiar velocities. Trials show that every factor of 3 increase in
the number of time steps results in a factor of 2 increase in precision.
With our set of parameters, a typical solution for given present
positions takes 1 hr with a DEC AXP workstation. The project began with
earlier and slower workstations, however, and approximately 2000 cpu
hours on various workstations have been expended on the calculations
here and on tests of robustness and accuracy.

\subsection{Scaling Laws}

We use two scaling relations. These results assume we can neglect
nongravitational forces and the cosmological constant $\Lambda$. They
can also be useful approximations when the universe is cosmologically
flat and $\Lambda\not= 0$.

The first relation is the simple result that the solutions for the
orbits of particles of specified masses, which follow proper physical
positions ${\bf r}(t)$ relative to the center of mass as functions of
proper time $t$ and end up at given present positions ${\bf r}(t_0)$,
depend only on the expansion time, $t_0$. This condition follows from
Birkoff's theorem: in a region with radius small compared to the Hubble
length the interactions of the particles may be described using
Newtonian mechanics (with a tidal field from the external mass
fluctuations). The background cosmological model serves only to set the
value of $t_0$. Peculiar velocities relative to the global expansion do
depend on $\Omega_0$ as well as $t_0$, of course, because the expansion
rate in the background cosmological model is determined by $H_0$.

The conclusion is that any cosmological model for which $H_0$ and
$\Omega_0$ imply a given $t_0$ predicts the same relation between mass
tracer distances and redshifts. In our action computations
we use a standard nominal value of the Hubble Constant,
and a range of values of $\Omega_0$
chosen to produce the wanted range of values of $t_0$. We have checked
that the action variational solutions based on different combinations of $H_0$
and $\Omega_0$ that give the same $t_0$ do give the same relations
between the mass tracer redshifts and distances.

A goal of this program is the analysis of the effect of large-scale mass
fluctuations on the reconstruction of the orbits of the nearer mass
tracers. Since the fluctuations on very large scales have small
density contrast
$\delta\rho /\rho$ it is efficient and reasonable to represent their
effect on the nearer mass tracers by an applied tidal gravitational
field that is a given function of comoving position and with amplitude
that grows with time according to linear perturbation theory. To the
extent that this treatment is a good approximation in the cosmological
model with the nominal values of $H_0$ and $\Omega_0$, it is also a good
approximation to the predicted distances in any other model with the
same value of $t_0$ and $\Lambda =0$.

The second relation is a gravity scaling law. Suppose we have a set of
comoving positions ${\bf x}_i(t)$ for all particles in a solution to an
N-body problem with pure gravitational interactions, and consider a
rescaling of proper times, comoving positions, and masses by the
equations

\begin{equation}
t'=t/\alpha ,\qquad {\bf x}'(t') = \beta {\bf x}(t)=\beta {\bf x}(\alpha t'),
		\qquad m'=\gamma m,
\label{eq_scale1}
\end{equation}
where $\alpha$, $\beta$, and $\gamma$ are constants. If the expansion
parameters in the original and scaled cosmological models satisfy the
relation
\begin{equation}
a'(t') = a(t) = a(\alpha t'),
\end{equation}
then the cosmological acceleration equation is
\begin{equation}
{\ddot a'\over a'} = -{4\over 3}\pi G\rho '
		= -{4\over 3}\pi G\rho \gamma\beta ^{-3}
		= \alpha ^2 {\ddot
a\over a}, \end{equation}
from which we see that the scaling parameters satisfy
\begin{equation}
\gamma = \alpha ^2\beta ^3 .
\label{eq_scale4}
\end{equation}
The Hubble constants in the original and scaled model satisfy $H_0' =
\alpha H_0$; the mass density scales as $\rho' = \alpha^2 \rho$; the density
parameters are unchanged: $\Omega_0' =\Omega_0$; the luminosity
derived from given apparent magnitudes scales as $L' = \beta^2 L$; and
the mass-to-light ratio scales as $M'/L' = \alpha^2 \beta M/L$. Finally,
one readily checks that the scaled orbits ${\bf x'}(t')$ satisfy the
same equation of motion as the original orbits ${\bf x}(t)$.

We use two applications of this scaling relation. First, the scales of
time and mass in our numerical results are set by observed redshifts and
the length scale of the luminosity-linewidth distances. If further progress in
the calibration of the length scale were to adjust the luminosity-linewidth
distances by the factor $\beta$, as in equation (6), then since the
velocities are not changed we would have $\alpha = \beta^{-1}$, meaning
expansion times and masses scale in proportion to the length scale.

Second, we can reduce the number of cases to compute with the action
principle, because the solutions for various mass-to-light ratios at a
given expansion time $t_0$ can be scaled to other values of $t_0$.  To
hold the redshifts fixed, we adjust $\alpha$ with $\beta$ =
$\alpha^{-1}$. In practice, though, only a limited range in $M/L$ is
interesting at each value of $t_0$ and it is often the case that the
range of values explored at one choice of $t_0$ do not rescale into the
interesting range at some other value of $t_0$. Rather, the benefit is
that we can calculate a coarse grid of parameter values and then use
scaling to fill in interesting parts of the grid at higher resolution.
As a check, we have used the rescaling to determine distances at
dramatically different values of $t_0$ and fixed velocities, and
compared the results to direct action variational computations at the
new values of $t_0$ and $M/L$. The results agree to a few percent with
the exception of a mass tracer in a triple-value region that jumped to a
very different orbit.

\section{Input Data}

The model predicts distances and masses of objects in a catalog of mass
tracer redshifts, positions on the sky, and luminosities, with the only
free parameters $t_0$ and $M/L$. This catalog is supplemented by initial
guesses of distances that serve to speed convergence of the computation
and discriminate among alternative possibilities when the solution is
not unique, as in triple valued regions. The influence of density
fluctuations on scales greater than 3000~\kms\ is modeled by a set of
positions of ``external masses.'' Finally, we match $M/L$ for the mass
tracers to the global mean value to determine $H_0$ and $\Omega_0$ in
terms of $t_0$ and $M/L$.  In this section we discuss the choices of all
these quantities.

\subsection{The Catalog of Mass Tracers at $\bf{cz<3000}$~\kms}

The mass tracer catalog is an extension of the one used in the linear
analysis by STP92 and given in Table 2 of TSP92. That earlier
description of the local distribution of galaxies and light was derived
from the {\it Nearby Galaxies} (NBG) catalog (Tully 1988$a$). The
`associations' defined by Tully (1987) constituted the mass tracers in
TSP92.  The current extension introduces three significant
modifications.

\smallskip\noindent 1. Galaxies in unbound associations are handled
differently.  It was appreciated that the linear analysis in STP92 was
inaccurate in high density regions so the mass tracers were grouped at
the low luminosity threshold of the `associations' in the terminology of
Tully (1987).  We can venture into higher density regions with the
present analysis. In principle, the action computation can be applied in
situations where the crossing time is short, although ambiguities arise
because we have no way to choose among the many extrema and saddle
points of the action. We want to avoid these complications in this
preliminary study, and have therefore chosen to group galaxies into mass
tracers that should not have experienced a close passage with any other
tracer. That is, we aim to construct the catalog so that the crossing
times between mass tracers is greater than $t_0$.

This condition is just the group identification criterion in Tully
(1987). In that earlier study, groups were identified on the basis of a
luminosity density threshold.  It was demonstrated that the chosen
threshold produced a group catalog with characteristic crossing times
less than or comparable to the age of the universe. Hence, for the
present study each Tully group is counted as  a single mass tracer.
Galaxies in Tully associations are below the luminosity density
threshold and enter the catalog as several mass tracers. To be rigorous,
we must note that there are some `associations' in the Tully (1987)
group analysis above the luminosity density threshold but that are
tidally sheared by an adjacent massive cluster.  Since such an entity
can have a short crossing time, the galaxies within it are consolidated
as a single mass tracer for the present study. Each mass tracer is
assigned the average velocity and angular position of its constituent
galaxies. In this paper the averages are number weighted; in future
papers we plan to consider the effect of luminosity weighting.

\smallskip\noindent 2. A substantial improvement to the low latitude
coverage is implemented with the merging of information from the IRAS
1.2 Jy survey (Fisher 1992, Fisher {\it et al.} 1992).  The improvements
are in two parts.  First, there is greater completion to low latitudes,
with uniform density maintained down to a latitude of $5^{\circ}$.
Second, we have a new procedure to deal with the equatorial band where
there is almost no information. The NBG catalog was meant to have uniform
all-sky coverage, within the constraints of galactic obscuration.  The
uniform, all-sky IRAS 1.2 Jy sample provides an important
supplement.

All 1304 IRAS 1.2 Jy sources with galactocentric velocities less than
3000~\kms\ are used in the updated database. Of these, 836 were already
in the NBG catalog and 468 are new.  The new objects are either at low
galactic latitudes or at high velocities ($> 1500$ \kms) where the NBG
catalog is known to be incomplete.  A new IRAS galaxy is integrated into
an existing mass tracer from the NBG catalog if it is within $2h^{-1}$
Mpc ($h = H_0/100$) of a tracer member in the appropriate sum of angular
position and velocity (where the differential radial distances are taken
to be $0.01 c\Delta z h^{-1}$ Mpc).  IRAS sources within $2h^{-1}$ Mpc
of each other and not incorporated in NBG tracers are grouped into a new
mass tracer, or isolated IRAS sources become themselves a new tracer.

The vast majority of IRAS sources have blue magnitudes in the {\it Third
Reference Catalogue of Bright Galaxies} (de Vaucouleurs {\it et al.}
1991). We apply the standard NBG catalog galactic and inclination
absorption adjustments to incorporate these new objects into our
database. In only 38 highly obscured cases are there no blue magnitudes.
In these cases, magnitudes are constructed from our IRAS count
normalization (1 IRAS count = $2 \times 10^{10} L_B \times f_{IRAS}$
where $f_{IRAS}$ is a distance incompletion factor which grows in a
slightly nonlinear way from unity at 1300~\kms\ to 2.5 at 3000~\kms).

The treatment of the galactic plane was unsatisfactory in the STP92
analysis and the new treatment to be described now represents a major
improvement of the current catalog.  The solid histogram in
Figure~\ref{catdis} illustrates the distribution of known galaxies in
the merged NBG catalog and IRAS sample, in sin~$|b|$ bins (where $b$ is
galactic latitude).  The histogram would be flat for a uniform galaxy
distribution and fair sampling.  It is seen that the characteristics of
our sample are consistent with this proposition for sin~$|b| > 0.09$
($|b| > 5^{\circ}$). The IRAS survey is impressively complete, at least
within the restricted velocity range $V_o< 3000$ \kms.

At $|b| < 5^{\circ}$ (9\% of the sky) our merged catalog is highly
incomplete.  We want to maintain a uniform luminosity density, so {\it
fake} mass points are created according to the following recipe.  All
real catalog mass tracers with $5\deg < b < 15\deg$ create a tracer with
the same distance and longitude but at latitude $b_{fake} = b_{real} -
10^{\circ}$. All real tracers with $-15\deg < b < -5\deg$ reflect to a
tracer with $b_{fake} = b_{real} + 10^{\circ}$. Each fake tracer is
assigned half the luminosity of its progenitor. By construction, the
luminosity in sin~$|b|$ bins at $|b|< 5\deg$  is consistent with that
at higher latitudes (dashed histogram in Figure~\ref{catdis}). Structure
is diluted by the halving of individual luminosities and the blending by
reflections from each side of the galactic plane.

\smallskip\noindent 3. Many new distance measurements are becoming
available, including quite a few to galaxies that are not in the NBG or
IRAS catalogs. In future studies we will use these data in the
goodness-of-fit measure, so the galaxies have been added to the mass
tracer catalog by the procedure used in introducing the IRAS 1.2 Jy
sources.  Again, these new galaxies all have redshifts $cz > 1500$~\kms,
where the NBG catalog is incomplete.

The mass tracer catalog sample has been corrected for residual
incompleteness as a function of distance following the procedure in
Tully (1988$b$),  but since the updated sample includes 28\% more
galaxies the correction is smaller.  The catalog is truncated at
$M_B^{b,i} = -16^m$.  At input distances less than 10 Mpc, there is no
correction for incompleteness.  At larger distances, the mass tracer
luminosities are multiplied by a correction factor that varies
non-linearly from $F_L^{new} = 1$ at 10 Mpc to $F_L^{new} = 2.4$ at 35
Mpc. In STP92 and TSP92 the maximum value of the correction factor was
$F_L^{old} = 3.0$.

There are 1323 mass tracers in the catalog in its present form. Of
these, 847 consist of a single galaxy. The largest tracers are the Virgo
Cluster, with 157 galaxies, and the Hydra~I and Centaurus clusters at
the edge of the sample with comparable luminosities.
Figure~\ref{lumbin} is a histogram of the number of tracers in bins of
log luminosity. Some of the tracers have quite low luminosities. In the
present computation we have applied a luminosity cutoff at $3 \times
10^9 L_\odot$ for tracers without distance measurements, but we keep all
tracers with measured distances. The resulting 1138 tracers are used in
the computations discussed in \S 5.

We also use a set of initial guesses for the mass tracer distances. For
most of the mass tracers the action solution is unique, but with reasonable
initial approximations to the distances fewer iterations are required to
generate good solutions. If there is more than one solution for a given
velocity then the computation will tend to find the solution with
distance nearest the input value. Hence, in regions where triple-value
zones are suspected, such as around the Virgo Cluster, we use measured
distances when available to discriminate between alternative input
distance choices.  For the other mass tracers the initial distance
estimates are in statistical agreement with the subset with measured
luminosity-linewidth distances, and we invoke continuity of filamentary
structures on the plane of the sky and in velocity. The same recipe was
used in our earlier linear analysis (STP92; TSP92). The initial trial
distance assignment for a new mass tracer from the IRAS sample is
the distance of the nearest neighboring NBG tracer multiplied by the
ratio of the redshifts of the new and NBG tracers.

The solution is tested against a subset of mass tracers with measured
distances. In the examples presented here, we use the TSP92 list of
distances, excluding nine mass tracers.  Five galaxies give distances
well removed from the predictions of all models (worse than $8 \sigma$).
These five distance measurements provide no model descrimination, likely
are bad data, and so are removed. One group is near the edge of the
sample with a few member galaxies beyond the velocity cutoff and must
therefore be removed.  Finally, three mass tracers are very near the
Virgo Cluster are thought to be caught in  the middle of the
triple-value region.  However, in models with low constant M/L, the
triple-value region does not go out far enough to include these
galaxies.  We will later detail how these objects indicate a greater
than nominal mass-to-light ratio for the Virgo Cluster.  This leaves us
with 289 galaxies providing velocity independent distance measures to
186 mass tracers.

\subsection{Beyond 3000 \kms}

Our redshift catalog of mass tracers represents the nearby region at
$cz<3000$~\kms\ which encompasses the historical Local Supercluster.
However, it is generally appreciated that this region is significantly
affected by mass fluctuations on larger scales (Lilje, Yahil, \& Jones
1986, Dressler {\it et al.} 1987, Strauss {\it et al.} 1992, STP92).
Here we describe a way to explore the effect.

It is too time consuming at present to compute the orbits of the mass
tracers at $cz > 3000$ \kms. Rather, we fix these distant tracers in
co-moving coordinates and scale the amplitude of their gravitational
field as a function of time according to linear perturbation theory. We
call the mass tracers treated this way `external masses'; they are
external to the volume with reconstructed orbits. As noted in \S 2.3, if
$\Lambda =0$ this procedure can be applied in a background cosmological
model with any combination of $H_0$ and $\Omega_0$ that yields the
wanted $t_0$.

There are two useful descriptions of the large-scale distributions of
objects: the IRAS and Abell catalogues. We use the QDOT IRAS sample
(Rowan-Robinson {\it et al.} 1990) and the sample of Abell clusters from
Lauer \& Postman (1994).  The two are similar enough that they both
induce a flow of the nearby mass tracers in the general direction of the
apex of the thermal cosmic background radiation dipole, and they both
generate a quadrupole tide along an axis running from Hydra-Centaurus to
Perseus. The method is illustrated in this paper by an external mass
model based mainly on the Lauer-Postman clusters.

Lauer \& Postman identified 128 clusters that they show are likely to be
a complete sample at high galactic latitudes and within 15,000~\kms. It
is known that the northern component (Abell 1958) is deficient in
richness class zero clusters compared with the southern component
(Abell, Corwin, \& Olowin 1989) but Lauer \& Postman argue that this
deficiency should not affect the sample at $z < 0.05$ because such
fairly nearby clusters are easily identified.  We use the relative
distance estimates that Lauer and Postman obtained for 120 of the clusters.

The distribution of the Lauer-Postman sample in latitude is shown in
Figure~\ref{lpdis}.  Incompletion apparently sets in rather abruptly at
$|b| = 24^{\circ}$ (sin~$|b| =0.4$). There should be 80 clusters
at $|b| < 24\deg$ to maintain a uniform density where only 8 are known.
This $48\deg$ equatorial gap is much greater than the $10^\circ$ gap in
the IRAS 1.2 Jy sample, so a different technique is used to fill in the
missing clusters. The IRAS 0.6 Jy 1-in-6 sparse-sampled (QDOT) survey
(Rowan-Robinson {\it et al.} 1990) has been provided to us in a binned
format (Kaiser {\it et al.} 1991). We are given an irregular grid with
the ratio of observed counts in cells to the counts expected for a
homogeneous distribution. We use the lowest latitude bins, $10\deg < |b|
< 32\deg$.  An `equivalent cluster count' is created with the product of
the normalized IRAS count in the cell times the bin volume size times a
bin normalization factor.  The bin normalization factor is chosen to get
the right overall number of sources in the equatorial sin~$|b|$ bins of
Fig.~\ref{lpdis}. Then the `equivalent cluster count' is assigned a
position, not at the center of the bin, but displaced to the edge of the
bin at $|b| = 10^{\circ}$, in order to fill the empty zone $-24\deg < b
< 24\deg$ with two bin strips with centers at $b = \pm 10\deg$.  As a
matter of detail, the `equivalent cluster counts' are fractional and are
smaller nearby, because the bin volumes are smaller, so we are not
placing any unduly large masses nearby. The assignment of these fake
clusters has some observational motivation, but they reflect the
distribution of IRAS galaxies rather than clusters, and the IRAS galaxy
distribution observed at $10\deg < |b| < 32\deg$ is taken to represent
the distribution in the offset interval $0\deg < |b| < 24\deg$.

The computations in \S 5 with external masses use this augmented cluster
sample. The net mass in the region occupied by the set of clusters and
specified by the choice of $\Omega_0$ is assigned equally to each of the
clusters and placed at the cluster centers. The gravitational field
produced by this mass distribution is tabulated as a function of
position in the region occupied by the test masses. The magnitude of the
gravitational field as a function of time is scaled from linear
perturbation theory, as discussed in \S 2.3. This prescription provides
the maximum tidal field that can reasonably be expected while keeping a
consistent relation between luminous density and mass density both
inside and outside the 3000~\kms\ volume.  The true tidal field strength
undoubtedly is somewhat less than what is given by this model; the
exploration of other cases is left for future papers.

\subsection{Luminosity Density Calibration}

The global mean luminosity density relates the mass-to-light ratio $M/L$
and the expansion time $t_0$, to the cosmological parameters $H_0$ and
$\Omega_0$.

Our mean luminosity density is based on the Automatic Plate Measurement
(APM) 1-in-20 redshift survey to $B_J = 17.15$ by Loveday {\it et al.}
(1992), at median redshift 15,000~\kms. This new calibration samples a
volume 30 times larger than the STP92 luminosity density calibration
based on the Center for Astrophysics redshift survey to $m_{\rm Zwicky}
= 14.5$ (Davis \& Huchra 1982), so there is a considerably better chance
now that we have a representative sample of the universe.  The new
calibration results in a 20\% decrease from the STP92 value for the
critical mass-to-light ratio in an Einstein-de~Sitter universe.

Our procedure follows Efstathiou, Ellis, \& Peterson (1988). A Schechter
(1976) luminosity function fit to the APM data provides three
parameters: the characteristic number density and faint end power law
index,
\begin{equation}
\phi^{\star} = 0.014 h^3 {\rm galaxies /Mpc^3},
\qquad \alpha = -0.97,
\end{equation}
and the characteristic luminosity,
\begin{equation}
L_B^{\star} = 1.28 h^{-2} \times 10^{10} L_{\odot},
	\qquad M_B^{\star} = -19.79 + 5 \log h.
\end{equation}
The determination of $L_B^{\star}$ in the $B_T$ system of the {\it
Revised Catalogue} (de Vaucouleurs {\it et al.} 1991) is based on a
transformation from the $B_J$ photographic magnitude system specified by
Efstathiou {\it et al.}, $B_T = B_J - 0.29$.

The mean luminosity density is
\begin{equation}
{\bar \ell_B} = \phi^{\star} L_B^{\star} \Gamma(\alpha+2),
\end{equation}
where $\Gamma$ is the gamma function. The above numbers give
\begin{equation}
{\bar\ell_B} = 1.76 h \times 10^8 L_{\odot} {\rm Mpc}^{-3}.
\end{equation}
Critical mass density ($\Omega_0=1$) is
\begin{equation} \rho_c = 2.78 h^2 \times 10^{11} M_{\odot} {\rm Mpc}^{-3},
\end{equation}
whence the global mass-to-light ratio is
\begin{equation}
\rho_c / {\bar \ell_B} = 1580 h M_{\odot} /L_{\odot}.
\end{equation}

This number is on the $B_T$ system but without absorption corrections,
whereas these corrections are applied to our mass tracer catalog. The
corrections amount in the mean to $0.20^m$ for inclination effects and
$0.04^m$ for galactic absorption in the domain of the APM survey with
$|b| > 40\deg$. Finally, a correction of 0.979 is applied because our
catalog has a luminosity cutoff at $-16$ mag.  These adjustments produce a 22\%
increase in ${\bar \ell_B}$, so in our system of corrected $B_T^{b,i}$
the global mass-to-light ratio is
\begin{equation}
\rho_c / {\bar\ell_B^{b,i}} = 1290 h M_{\odot} /L_{\odot}. \label{eq_critml}
\end{equation}

We have a check from  the IRAS 1.9 Jy sample.  The Yahil {\it et al.}
(1991)  estimate of the mean number density of sources is ${\bar n_{\rm
IRAS}} = 0.045 h^3 {\rm Mpc}^{-3}$, at their luminosity cutoff. Then
the critical mass per IRAS source is
\begin{equation}
\rho_c / {\bar n_{IRAS}} = 6.2 h^{-1} \times 10^{12}M_{\odot} .
\end{equation} For the
IRAS sources within 3000 \kms\ with measured $B_T^{b,i}$ magnitudes the
optical luminosity per source is
\begin{equation}
{\bar \ell_B^{b,i}} /{\bar n_{IRAS}} = 4.7 h^{-2} \times 10^9 L_{\odot} .
\end{equation}
The ratio of these two numbers,
\begin{equation} \rho_c / {\bar\ell_B^{b,i}}\simeq 1300 h M_{\odot}/L_{\odot},
\end{equation}
is in excellent agrement with the APM normalization in
equation~(\ref{eq_critml}).

It is difficult to measure the mean luminosity density and evaluate the
potential for systematic errors; our soft estimate is that
equation~(\ref{eq_critml}) could be in error by 20\%.

\section{Statistical Measures and Biases}

We discuss here the procedure to obtain an unbiased distance measure and
the $\chi^2$ statistic we use to compare the measured distances with
those predicted by a numerical solution for the orbits of the mass
tracers. Here is a one sentence summary of our procedures.  The
regression on {\it linewidths} in a faint magnitude-limited calibrator
sample provides the slope required to measure unbiased distances to
individual galaxies in the field; errors are symmetric in the {\it
logarithm} so moduli are measured; and volume error-scatter effects are
minimized by evaluating {\it distances at observed velocities} rather
than velocities at observed distances.  Details follow.

\subsection{The Distance Estimator}

Several problems have been termed Malmquist bias (eg, Teerikorpi 1984,
Lynden-Bell {\it et al.} 1988, Tully 1989, Willick 1994). One arises
because of the magnitude or diameter cutoffs that almost always limit
samples (Malmquist 1922); another because distance errors tend to
preferentially throw galaxies to smaller distances because more distant
galaxies are more numerous (Malmquist 1920); yet another because
distance errors for the galaxies in a concentration masquerade as a
converging peculiar velocity field. A common practice in the last decade
has been to deal with such problems by the introduction of correction
terms. This is a difficult art, however, in which even the sign of the
correction can be in doubt, so we adopt and strongly advocate a
different procedure in which one attempts to {\it neutralize} biases by
using estimators that are likely to be close to unbiased, rather than
seeking to correct for anticipated biases.

The two steps in the establishment of luminosity--linewidth distances
with the method of Tully \& Fisher (1977) are the calibration of the
relation and the method of application.  In our procedure the first step
is to find the mean of the linewidth parameter as a function of the
galaxy absolute magnitude rather than the mean absolute magnitude as a
function of linewidth, that is, we use the so-called inverse rather than
direct calibration.  The alternative is liable to the 1922 Malmquist
bias. In the second step we use the observed galaxy apparent magnitude
and linewidth to predict the distance modulus, which is compared to the
model prediction. Since completeness is likely to be insensitive to
linewidth at a given magnitude, our estimator of the distance modulus is
unbiased, as desired. An alternative approach uses the linewidth,
apparent magnitude, and redshift to estimate the galaxy peculiar
velocity.  This latter approach is subject to the 1920 Malmquist bias.

To make these points more explicit we illustrate our procedure in the
notation of Willick (1994) as follows. The linewidth parameter is
\begin{equation}
	\eta = log W_R^i - 2.5 ,
\end{equation}
where $W_R^i$ is an inclination-corrected HI profile linewidth
observable related to twice the maximum rotation velocity (Tully \&
Fouqu\'e 1985). We assume that an absolute magnitude characteristic of
galaxies with linewidth parameter $\eta$ is given by the linear
luminosity-linewidth relation
\begin{equation}
	M = a - b\eta,
\end{equation}
where $a$ and $b$ are constants.

Suppose the joint probability distribution in absolute magnitude and
line width parameter in a fair volume-limited sample of galaxies is
\begin{equation}
	dN = e^{-(\eta - a/b + M/b)^2/2\sigma (M){}^2}F(M)dMd\eta .
\label{eq_dN}
\end{equation}
The rms spread $\sigma (M)$ in $\eta$ among the galaxies with given
absolute magnitude $M$ may depend on the value of $M$, and the function
$F(M)$ takes account of the luminosity function and normalization.
Equation~(\ref{eq_dN}) says the average value of $\eta$ in a fair sample
of galaxies with given $M$ is
\begin{equation}
	\langle\eta\rangle _M = (a-M)/b .
\label{eq_eta}
\end{equation}
To calibrate the luminosity-linewidth relation expressed in this form
one requires a fair sample of galaxies at known distances. To this end
we use complete samples of galaxies in one or several clusters
(specified for this study at the end of the section). With this
calibration sample, a regression that minimizes the mean square scatter
in $\eta$ as a function of $M$ yields the parameters $a$ and $b$ defined
in equations~(\ref{eq_dN}) and (\ref{eq_eta}).

The estimator of the distance modulus for a galaxy observed to have
apparent magnitude $m$ and line width parameter $\eta$ is
\begin{equation}
\mu _e = m - a + b\eta = \mu + M - a + b\eta .
\end{equation}
The last expression follows if the true distance modulus of the galaxy
is $\mu$. As Willick (1994) and others have clearly emphasized, this
estimator is biased, when there is incompletion, in the sense that the
expectation value of $\mu _e$ at given $m$ and $\eta$ is not equal to
the true value $\mu$. Furthermore, we cannot determine the amount of
bias unless we have good models for $\sigma (M)$ and $F(M)$. But our
dynamical analysis is not sensitive to the value of $\eta$, and we can
assume that the selection of galaxies at given apparent magnitude $m$ is
not very sensitive to $\eta$. This independence from $\eta$ means that
for the purpose of the analysis of the peculiar velocity field we need
the expectation value of our estimator for given $m$:
\begin{equation}
\langle\mu _e\rangle = \int dMd\eta F(M)(\mu + M - a + b\eta ) e^{-(\eta
- a/b + M/b)^2/2\sigma (M){}^2} .
\end{equation}
It follows from equation~(\ref{eq_eta}) that
\begin{equation}
\langle\mu _e\rangle = \mu .
\end{equation}
That is, the estimator is unbiased. This is the quantity used in
equation (28) to measure the goodness-of-fit of the model to the
observations.

We could reduce the statistical scatter in the measure in equation~(28)
if we corrected equation (24) to the expected value of the distance
modulus $\mu$ at given $\eta$ and $m$. The procedure does not seem
advisable at our present state of knowledge, however, because an
inadequate model for the correction would introduce a systematic error
in the distance modulus.

Several features of this procedure might be noted and contrasted with
other common practices. First, it is essential to our approach that the
galaxies to which the estimator is applied are {\it statistically
similar in properties to the calibrators}, with similar measurement
uncertainties. These properties do not have to be specified beyond the
definition of the slope and zero-point of the regression to the
calibrators, however. The completeness of the calibration sample in
general will depend on the absolute magnitude $M$, but that
requirement introduces
no bias as long as the sample is large enough to determine the mean
value of $\eta$ as a function of $M$. As Willick (1994) points out, our
calibration may be biased if galaxies are selected by diameter or by
magnitude in a different passband from that of the distance estimator
correlation, as almost inevitably is the case, and if linewidths are
correlated with galaxy surface brightnesses or colors. However,
Willick's realistic simulations indicate that this bias is negligible in
our sample, limited as we are to $cz < 3000$ \kms.

Our procedure often is called the inverse Tully-Fisher relation, and the
expectation value of the absolute magnitude given $\eta$ is called the
forward relation.\footnote{The terminology is curious, because RBT has
never advocated the use of what has come to be called the forward
relation.}  The forward relation suffers the Malmquist (1922) bias in
the domain where the scatter in the correlation overlaps the sample
selection magnitude limit.  In this domain, $\langle\mu _e\rangle < \mu
$. The problem can be avoided by truncating the sample in linewidth or
something correlated with linewidth (Bottinelli {\it et al.} 1986, Rood
\& Williams 1993, Lu, Salpeter, \& Hoffman 1994), but at the cost of a
substantial loss in sample size. If instead one attempts to correct for
the bias (Teerikorpi 1984, Willick 1994) one must be able to specify the
characteristics of the distance-dependent correlations.  In particular
one must know the convolution of intrinsic and observational dispersions
and the often subtle effects of sample selection. In our approach, on
the other hand, {\it the onus of sample completion is transferred to the
calibration}. As long as we are not inhibiting the sample with the
measurements of linewidth, statistically unbiased distances are being
measured over the full range of calibrated luminosities, whatever the
degree of completion of the sample under study as a function of
magnitude.  We can use all observations of non-pathological galaxies and
make a statistically unbiased distance estimate for each one.

The present preliminary study uses the calibration by Pierce \& Tully
(1988, 1992) based on a deep sample of Ursa Major Cluster galaxies. In
future papers in this series, the calibration will be based on a more
extensive 3-cluster sample and applied to a much larger number of
objects dispersed around the Local Supercluster.

\subsection{Volume Malmquist Bias}

Our strategy in building the dynamical model also differs from the more
familiar practice in that our dynamical analysis is based on a redshift
map rather than a map of the mass distribution in real space. The former
approach, which has been taken also by Nusser \&\ Davis (1994), has not
been as widely discussed, so its hazards surely are not yet as well
explored as the real space construction. We can note a very apparent
advantage, however. Measurement errors distort either a real space or
redshift map (Malmquist 1920; Lynden-Bell {\it et al.} 1988). For
example, in a real-space map, if galaxies are uniformly distributed then
the volume effect causes more galaxies to be thrown forward by errors
than thrown back.  With available distance measurements the error in
distance of a single galaxy is 15 to 20\%. These large percentage errors
cause a considerable distortion that grows linearly with increasing
distance. By contrast, the scatter in our redshift catalog of mass
tracers due to the small-scale motions we cannot model are expected to
be 100 to 200~\kms, a 3 to  7\% effect at 3000 \kms\ and constant in
absolute terms with increasing distance.

\subsection{Velocity-Limited Volume Bias} There is one significant
deficiency to a velocity-limited sample as opposed to a distance-limited
sample: the boundary of the sample volume is not spherical but rather
depends on the peculiar velocity field we are computing.  In the context
of these calculations, missing regions of a sphere subtending the
sampled volume effectively have density contrast $\delta \rho / {\bar
\rho} = -1$, and this feeds back to cause accelerations that enhance the
flattening. The result is that in early times the missing regions are
inappropriately filled by mass tracers from the sample.

Distortion of the boundary arises from the peculiar velocities of the
Local Group and at the sample boundary. The effect of Local Group
peculiar velocity $V_{LG}$ is to shift the edge of the sample at cutoff
redshift $cz_{max}$ to the distance \begin{equation}H_0 R = cz_{max} +
V_{LG}\cos\phi , \end{equation} where $\phi$ is the angle away from the
direction of motion. Perpendicular to the peculiar velocity the sample
boundary is at the appropriate distance, but in the direction of the
Local Group motion it is  $V_{LG}/H_0$ too distant, and behind the
motion it is too near by the same amount.  The off-centering of the
Local Group is of no consequence, because we compute in center of mass
coordinates. The problem is that the sample volume is deformed. Examples
of possible shapes are presented in Figure~\ref{sampleshape}. One sees
that for values of $V_{LG}/cz_{max} < 0.5$ the departure from an offset
sphere is quite small, so we do not have a significant problem, but we
would have if we had chosen a cutoff redshift at $cz_{max}=1000$~\kms.

As an aside, there are some curious effects with a sample boundary
deformed by local motion.  The sample volume is enlarged from the zero
peculiar velocity case by the factor $(1 + r^2)$, where $r =
V_{LG}/cz_{max}$. An observer looking into the two hemispheres, in the
direction of motion and away, sees volume effects that could be
substantial. In the hemisphere of the motion, the volume is increased
from the expected hemisphere volume by $(1 + 3r/2 + r^2 + r^3/4)$. In
the other hemisphere, the volume  is decreased by a factor of $(1 - 3r/2
+ r^2 - r^3/4)$.  This deformation can be important when determining
luminosity functions and selection functions from galaxies in the Local
Supercluster region. We have yet another reason to build an accurate
model of the Local Supercluster.

A second cause of nonsphericity is the large-scale peculiar flow at the
boundary. A concrete example of this problem arises with the
circumstance that the center of mass of the Local Supercluster is
distinctly offset within the  3000 \kms\ shell, with three times the
light in the north galactic hemisphere as in the south. The inward flow
of the north galactic pole boundary is greater than at the boundary in
the opposite direction.  Therefore, the redshift cutoff of the sample
refers to a greater distance in the Virgo Cluster direction, which is
near the north galactic pole.  Likewise, the tidal fields from large
high density regions outside the sample volume  cause distortions.  In
some cases, the triple-value region of an external mass will be near the
3000 \kms\ boundary so that the mass tracers beyond the external mass
are thrown into the redshift range of our sample, and tracers which
otherwise would be included are thrown out of the sample. This situation
appears to actually be the case in the region near the Hydra and
Centaurus clusters.  However, in the current preliminary analysis very
few galaxies with distance estimates are close to the boundary near the
Hydra-Centaurus region, so inaccuracy in the detailed mass structure in
that region does not significantly affect the tests.

The problems of a nonspherical boundary are greatly reduced by including
the mass structures beyond the redshift cutoff.  It does not matter if
the mass tracers used in the dynamical computation are subject to an
aspherical cutoff in distance if one can place external masses with
sufficient accuracy to generate their proper contribution to the nearby
peculiar flow. Placing objects with redshift $cz> 3000$~\kms\ at their
Hubble flow distances should be satisfactory for the present purpose.
However, the 3000 \kms\ discontinuity in the present models may be the
greatest weakness of the current work and we hope to give this issue
closer attention in future papers.

\subsection{Test of the Fit of the Model to the Observations}

Our measure of the degree of consistency of the model prediction and the
observations is the reduced $\chi^2$ statistic \begin{equation} \chi^2
= \sum_{i=1}^N n_i \left({\mu_{m,i} - \mu_{TF,i} 	\over
0.4}\right) ^2 / \sum_{i=1}^N n_i . \label{eq_chi2} \end{equation} Mass
tracer $i$ has $n_i$ galaxies with distance measurements, the mean value
of the measured distance moduli for mass tracer $i$ is $\mu_{TF,i}$, and
the model prediction for the distance modulus of the tracer is
$\mu_{m,i}$. We have adopted a conservative estimate of the standard
deviation in the measurement of an individual distance modulus, $\sigma
(\mu ) = 0.4$~mag. In the sample used in \S 5, 186 mass tracers contain
one or more galaxies with measured distances.

We use the distance modulus $\mu$ rather than the distance because the
error in the luminosity-linewidth (Tully-Fisher) distance is symmetric
in $\mu$. We show in the next section that the scatter in $\mu_{m,i} -
\mu_{TF,i}$ is not very sensitive to the redshift of the tracer and that
the scatter is close to what is expected from the measurement error
$\sigma (\mu )$. If, as this result indicates, $\chi^2$ is dominated by
the distance measurement error, then because the number of degrees of
freedom is large we can approximate the standard deviation in the value
of $\chi^2$ as

\begin{equation}
\sigma (\chi^2) = \left( 2/ \sum_{i=1}^N n_i \right) ^{1/2} = 0.08,
\label{eq_sigchi}
\end{equation}
for the 289 galaxies with distance measurements used here.

A more commonly used measure of the goodness-of-fit is the mean square
value of the difference $V_{obs} - V_{model}$ between measured and
predicted velocities at a measured distance (Strauss 1989, STP92).
Evaluation of a $\chi _V^2$ statistic in this case requires the standard
deviation of $V_{obs} - V_{model}$, which is the sum in quadrature of
the measurement and model uncertainties. Since the latter depends on the
value of the density parameter, an error in modeling the standard
deviation of $V_{model}$ as a function of $\Omega_0$ can produce a
systematic error in the value of $\Omega_0$ at the nominal minimum of
$\chi _V^2$.  The important point here is that the prediction of
velocities as a function of position may be much more density sensitive,
hence subject to bias, than is the prediction of positions as a function
of velocity.

\section{Numerical Results}

The computations presented here use the redshift catalog of 1138 mass
tracers at $cz<3000$~\kms\ described in \S 3.1 and  289
luminosity-linewidth distance measurements listed in TSP92. We compare
two cases, where alternatively the mass fluctuations external to the
redshift catalog are neglected or are modeled by the distribution of
Lauer-Postman Abell clusters described in \S 3.2.

A specific example of the orbit solutions is shown in
Figure~\ref{tadpoles}.  Only galaxies concentrated to the supergalactic
plane are illustrated here.  The diamonds denote positions today.  The
orbits of selected objects are indicated by the curved lines (comoving
coordinates with respect to the sample center of mass).  The Local
Group, at the origin, the Virgo Cluster, and Ursa Major are identified
by very thick lines. The other objects with identified orbits are all
mass tracers associated with the Coma-Sculptor Cloud in the NBG
nomenclature. This specific example is an illustration of a reasonable
model but it does not take into account the effects of masses beyond
3000 \kms.  If distant masses are considered then a dominant
characteristic of the orbits is coherent streaming toward the minus SGX
direction.

\subsection{Contour Maps of $\chi^2$}

We have computed numerical action principle solutions at about 20 points
in the (M/L, $t_0$) plane and applied rescaling (\S 2.3) to double the
resolution of this grid. The values of reduced $\chi^2$ at each grid point are
interpolated by a cubic polynomial fit at each $t_0$ value followed by a
cubic spline fit at each M/L value to produce the contour maps of the
reduced $\chi^2$ statistic (equation~[\ref{eq_chi2}]) shown in
Figure~\ref{chi2map}. In panel $a$ there are no external mass
contributions. Panel b shows the effect of the external mass
distribution as modeled by the Lauer-Postman sample of Abell clusters.
The contour lines are placed at one standard deviation intervals
(equation~[\ref{eq_sigchi}]). The minimum values of $\chi^2$ in the
maps are close to unity: $\chi^2 = 1.29$ for no external masses and
$\chi^2 = 1.17$ with the external mass model. These values are
consistent with the proposition that the scatter between the model
predictions and the observations is dominated by the distance
measurement errors with some contribution from deficiencies in the
model.

The axes of the figure are the free parameters in the computation of
$\chi^2$: the mass-to-light ratio $M/L$ of the mass tracers and the
expansion time $t_0$ of the cosmological model. If $M/L$ in the region
modeled by these mass tracers is consistent with the universal value,
then we can use the measured global value, $M/L = 1290h \Omega_0$
(equation~[\ref{eq_critml}]), to compute the cosmological parameters
$H_0$ and $\Omega_0$ as functions of $t_0$ and $M/L$. The dotted and
dotted-dashed lines in Figure~\ref{chi2map} are loci of constant $H_0$
and $\Omega_0$ respectively.  The constraints placed on $\Omega_0$ and
the mass-to-light ratio, in the end, is quite similar to those from the
linear analyis of STP92 as can be seen by comparing Figure 4 of that
paper with this figure.

If the mass tracer catalog described a homogeneous and homogeneously
expanding mass distribution then the valley of the minimum of $\chi^2$
would have a flat bottom along a line of constant rate of expansion of
the system of mass tracers. The minimum of $\chi^2$ in the valley
and the twisting off of the constant rate of expansion result
from the density fluctuations in the catalog. For small expansion ages,
the masses become quite large along a line of constant rate of
expansion.  To avoid excessive peculiar motions the best models must
shift to lower mass-to-light ratios and thus to higher values of $H_0$.

For the case where external masses are neglected, Figure~\ref{minchi}a
displays the minimum value $\chi^2$ as a function of $t_0$ (dashed
curve, right scale) and the value of M/L at the minimum $\chi^2$ as a
function of $t_0$ (solid curve, left scale). The latter is usefully
approximated as
\begin{equation}
	M/L = 1100 \exp[(7-t_0)/2.2] - 100.
\end{equation}
At one standard deviation from the minimum of $\chi^2$ the parameters are
\begin{equation}
	9.0 < t_0 < 11.5 \hbox{ Gyr},\qquad \Omega_0= 0.17 \pm 0.10 .
\end{equation}

The effect of the model for the external masses is shown in
Figure~\ref{minchi}b. The minimum of $\chi^2$ drops by about one
standard deviation, the valley shifts to an empirical fit of
\begin{equation}
	M/L = 900 \exp[(7-t_0)/2] - 50,
\end{equation}
and the one standard deviation range shifts to
\begin{equation}
	9.0 < t_0 < 11.0 \hbox{ Gyr},\qquad \Omega_0= 0.15 \pm 0.08 .
\end{equation}
When the external mass model is applied, the curvature of the floor of
the $\chi^2$ valley is larger, so high and low values of $M/L$ are more
strongly excluded.

If $t_0$ is required to be no less than 11~Gyr, which seems to be a very
conservative bound from stellar evolution and radioactive decay ages,
then the solutions require that the density parameter is not much larger
than about $\Omega_0= 0.1$, because the Hubble parameter is not less
than about 80 km s$^{-1}$ Mpc$^{-1}$. The expansion time issue is
discussed further in \S 5.5.

In our model the major part of the tidal field of the external masses on
the mass tracers comes from the nearest superclusters, Hydra-Centaurus
and Perseus-Pisces at $cz =3500$ and 5500~\kms. These two superclusters
are situated on opposite sides of the sky near the galactic plane.  The
Virgo Cluster is only $15\deg$ from the north galactic pole and
therefore is nearly perpendicular to the axis connecting the nearest
adjacent superclusters.  The Local Group thus experiences tidal
contraction toward Virgo. To counter this effect and preserve a
reasonable value for the peculiar motion of the Local Group and
neighboring groups toward the center of the Local Supercluster one must
either reduce the mass density in the Virgo region or reduce the age of
the universe. We interpret this tidal effect as the cause of the tighter
constraint on $M/L$ and $\Omega_0$ when the external mass model is
applied.

\subsection{Observed and Model Distances}

Figures~\ref{delmu_cz_noex}, \ref{delmu_cz}, and \ref{delmu_cz2}  show
scatter plots of the residual differences $\Delta\mu$ between observed
and model distance moduli as a function of the redshift of the mass
tracer. Figure~\ref{delmu_cz_noex} presents a case near the minimum
$\chi^2$ for no external masses. Here and along the $\chi^2$ minimum
valley the residuals of $\Delta\mu$ consistently trend downward with
increasing redshift $cz$. The trend is reduced if the mass-to-light ratio
is increased, but at the expense of making the model distances
systematically larger than the observations. We expected (Shaya,
Peebles, \& Tully 1994) that the problem would be relieved by the
addition of external masses: it was reasoned that the correlation of
$\Delta\mu$ with $cz$ is an indication of excess overdensity at large
Virgocentric distances when the density is about right in the local
neighborhood. As we noted in the last subsection, the external masses
cause tidal compression between us and the Virgo cluster which allows a
lower local mass density. In Figure~\ref{delmu_cz}, for the external
masses near the $\chi^2$ minimum at $t_0 = 9.8$ Gyr with $M/L = 160$, the
trend of $\Delta\mu$ with $cz$ is reduced, consistent with this
reasoning. Figure~\ref{delmu_cz2} shows the external mass case with
$t_0=8.0$ Gyr with $M/L=500$, which is in the $\chi^2$ valley but $\sim 4
\sigma$ from the best value. One is struck by the subtlety of the
differences between the scatter plots in these large and small
$\Omega_0$ cases.   However, when these plots are placed one over the
other, one sees the scatter is consistently greater in
Figure~\ref{delmu_cz2}, as the reduced $\chi^2$ indicates.

\subsection{Observed and Predicted Peculiar Velocities}

Figure~\ref{um_u0} shows a scatter plot of observed peculiar velocities
$U_{\rm obs}=cz - H_0 D_{TF}$ for the mass tracers with
luminosity-linewidth distances against the model peculiar velocities
$U_{\rm m} = cz - H_0 D_{m}$, where $D_m$ is the model prediction of the
distance. The model is near the $\chi^2$ minimum with external masses
($M/L = 160$ and $t_0 = 9.8$). Although our analysis is independent of
$H_0$, we must choose a specific value to derive peculiar motions, so
we have chosen
90 km/s/Mpc to be consistent with the APM survey luminosity density
normalization and this mass-to-light
ratio.  The correlation coefficient is 0.42 and the probability of
obtaining a higher value with 186 data points from a random distribution
is well below 1\%.  However, a high correlation coefficient is not
necessarily proof of a true correlation because if one chooses a
sufficiently low value for $H_0$ one is guaranteed a strong
correlation (the  correlation coefficient goes to unity as $H_0$ goes
to zero).  To
the extent that one believes 90 is not a low value for $H_0$, one can
believe that this figure indicates that peculiar velocities actually
have something to do with gravitational influences acting over time.

	A regression based on equal errors on the two axes would result
in a slope significantly greater than unity and would imply that the
model considerably underestimates the peculiar velocities. This analysis
could be very misleading, however. The errors in the observed peculiar
velocities are dominated by the errors in the distance measurements,
which amount to 600~\kms\ at the edge of the sample, and are comparable
to or larger than the expected true peculiar velocities. The numerical
action model is expected to do better because it is, at least, likely to
get the sign of the peculiar velocity correct most of the time in
predicting that galaxies tend to fall in toward regions of neighboring high
galaxy density. Consistent with this argument, we have the evidence from
the preceeding subsection that the distance modulus residuals are
dominated by the errors in the measured distance moduli. It follows that
a less biased regression would minimize the errors in the observed
quantities as a function of the model prediction. As indicated in the
figure, the slope of this regression fit is close to unity.

Figure~\ref{dUvsU} shows the data in Fig.~\ref{um_u0} another way, as a
scatter plot of observed minus model peculiar velocity as a function of
the predicted peculiar velocity of the mass tracer. This plot may be a
useful way to identify systematic errors in the model. In particular, a
few large deviations, including three groups excluded from the analysis
which would lie off the plot region, are in the direction of the infall
region around the Virgo Cluster. These mass tracers are not placed near
the positions we expected based on our earlier work on infall into the
Virgo Cluster (Tully \& Shaya 1984), but they would be if we increased
the mass-to-light ratio of the Virgo Cluster relative to the mass
tracers in less dense regions.  We have run a few trials in which a
larger value of $M/L$ is assigned to the mass tracers associated with
the richest clusters. The indications are that the resulting increased
sizes of the triple-value regions creates alternative positions for the
mass tracers that are in better agreement with the measured distances.
We hope to present a detailed analysis of this issue in a separate
paper.

The positive correlation of observed and peculiar velocities in
Figure~\ref{um_u0} is evidence that the peculiar velocities are being
observed in a statistical sense, and the correlation coefficient is
about right if, as we have argued, the scatter in the measured peculiar
velocities is dominated by the distance errors. This comparison of
observed and predicted peculiar velocities is commonly used as a test of
models for the deviations from pure Hubble flow, and indeed was the main
statistical measure in the early stages of our program. The
difficulty of dealing with the mix of measurement errors and the
$\Omega_0$-dependence of model velocities has led us to prefer
the rms difference of predicted and observed distance moduli.

\subsection{The Local Density Contrast}

As the mass-to-light ratio is increased, the model distances generally
increase to maintain the required match with the observed redshifts and
avoid excess infall velocities into the Virgo Cluster and the Local
Supercluster in general. The result is a decrease in the local
luminosity density relative to the global mean. In the same way, an
increase in $t_0$ increases the distances because increased expansion
time with given masses also tends to increase the inward peculiar
velocities. The valley of the  minimum of $\chi^2$ is established by the
trade-off between an increase in $t_0$ and a decrease in the
mass-to-light ratio: densities and distances remain close to optimal
when compared with the measurements of distances. Figure~\ref{density}
gives examples of these trends in terms of the effective luminosity
density profile.  The origin of this figure is at the Local Group  and
the model output distances are used to specify the mean luminous density
within a given distance from the Local Group.  The horizontal lines show
the APM global mean luminosity density. The solid line shows a model at
the minimum of $\chi^2$, with $H_0$ = 85, $\Omega_0$ = 0.2, M/L = 200,
and no external masses. Here the region remains slightly overdense to
the edge of the sample just beyond 30 Mpc. In the case with the lowest
density contrast (long dashes), the value of M/L has been increased to
800 while $t_0$ was unchanged. In the model with the highest density
contrast  (triple-dot-dash) the mass-to-light ratio has been lowered to
$M/L=45$, $t_0$ was again unchanged.  The other two cases are on the
minimum valley of $\chi^2$, with $\Omega_0 = 0.044$ and M/L = 45 (dots)
and $\Omega_0 = 0.42$ and M/L = 475  (short dashes). They show, as
expected, similar density profiles with only minor readjustments in
distances of specific features.

\subsection{The Expansion Time Problem}

The fit of our model to the distance measurements requires an expansion
time $t_0$ near 10~Gyr, consistent with a Hubble constant in the range
of 80 to 90 km s$^{-1}$ Mpc$^{-1}$. Since this value for the expansion
time seems uncomfortably short, we mention three possibilities often
mentioned to increase the age. First, physical lengths and times in our
computation are tied to the distances assigned to the
luminosity-linewidth calibrators. If these distances were multiplied by
the factor $\alpha$ it would multiply the expansion time by the same
factor (eqs.~[\ref{eq_scale1}] to [\ref{eq_scale4}]). The leeway for
adjustment here seems to be only about 10 -- 15\%, however, because our
distance scale is consistent with other recent measurements (Jacoby {\it
et al.} 1992; Freedman {\it et al.} 1995).  Second, the age difficulty
has motivated consideration that we live in a very large low density
region, which would make the global value of $H_0$ significantly lower
than the locally measured effective value (Turner, Cen, \& Ostriker
1992).  However, the parameter we directly measure is $t_0$, not $H_0$,
so the normalization of $H_0$ does not affect our ages.  Third, if the
density parameter $\Omega_0$ is less than unity a cosmological constant
that makes the universe cosmologically flat increases the expansion time
relative to an open universe with the same value of $\Omega_0$.

In a preliminary exploration of this possibility we computed a grid of
solutions with nonzero $\Lambda$ by incorporating the appropriate
variant of equation~(\ref{eq_worldmod}). The values of $\chi^2$ as a
function of $M/L$ and $\Omega_0$ are remarkably similar to the open
model case. That is, there is the indication that our constraints on
$\Omega $ and $M/L$ would be little changed if one used a cosmological
constant to increase the age of the universe and make the universe flat
and consistent with the standard picture for inflation.  However, if
$\Omega_0 = 0.2 (0.1)$ then the age of the universe is only increased by
26\% (41\%).   If the distance scale zero-point is unchanged, flat
universe models with $\Omega_0 > 0.1$ have $t_0 < 15$ Gyr.

\subsection{Motion of the Local Group}

The peculiar velocity of the Local Group that arises in the models may
be compared to the motion of the Local Group deduced from the dipole of
the cosmic microwave background. The large plus sign in
Figure~\ref{lgvel} shows the motion of the Local Group with respect to
the cosmic microwave background reference frame in supergalactic
coordinates. The solid curve shows the motion of the Local Group in our
model for solutions in which the external masses are neglected, and the
dashed curve shows the effect of adding the external masses. The
parameters are in the valley of the minimum of $\chi^2$; the solutions
are labeled by the expansion time $t_0$.   Without external masses, one
sees the influence of the Local Supercluster on the peculiar motion of
the Local Group.  For one of the best models with external masses, where
$t_0$ = 9.2 Gyr, and M/L = 250, the peculiar velocity of the Local Group
in supergalactic coordinates is (--346,96,--744) = 763 \kms\ to right
ascension $\alpha = 8^h15^m$ and declination $\delta  = -33.5^\circ$,
while the microwave background dipole velocity is (--318,315,--348) =
567 \kms\ to $\alpha = 10^h33^m$, $\delta = -24.0^\circ$ (Bennett {\it
et al.} 1994 with solar motion 250 \kms\ to $l=90$, $b=0$). It is
actually reassuring that the model amplitude is greater than the actual
amplitude because the maximal tidal field model was used.  One possible
prescription to correct the model would be to; 1) decrease the effective
masses of the clusters by about 30\%, 2) double the mass for the Virgo
Cluster to increase the SGY component, and 3) decrease the influence of
the Pisces and Perseus Clusters by about 25\% to increase the negative
SGX component and decrease the negative SGZ component.  Of course, many
other prescriptions are possible.  It is also reasonable that
inhomogeneities beyond 15,000 \kms\  contribute significantly to the
Local Group motion. With our present study, we have only
made a preliminary, naive foray into the problem of large scale
contributions and the microwave background constraints.

\section{Discussion}

We comment here on three issues we hope to be able to address in more
detail in future papers in this series.

\subsection{Biases in Measurements and Models}

Our aim in this analysis of the departures from pure Hubble flow has
been to set up the methodology of the astronomical measurements, the
model computations, and the statistics that compare the two so that we
have made the biases tolerably small and not in need of corrections.
The dynamical model predicts the distances of mass tracers given their
angular positions, redshifts, apparent magnitudes, mass-to-light ratios,
and the expansion time of the cosmological model. The predicted distance
moduli $\mu_m$ are compared to astronomical measurements so arranged
that the bias in the measured values $\mu_{TF}$ is likely to be small.
Perhaps the major issue here is whether the cluster members used in the
calibration are indeed statistically similar to the galaxies in groups
and associations.  A test which provides some confirmation is the
reasonable agreement in the form of the calibration when separate
clusters are evaluated.  This consistency will be discussed in
connection with the 3-cluster calibration that will be used in
future work.  Another test is the consistency of the residuals $\mu
_{TF}-\mu _m$ in clusters and the field. The indication from
Figure~\ref{delmu_cz} is that the scatter in $\mu _{TF}-\mu _m$ is about
what would be expected from measurement errors that are insensitive to
distance or environment. There is a systematic trend of the residuals
with distance which we are inclined to attribute to two effects, the
gravitational field of large-scale mass fluctuations and a higher value
of $M/L$ in the neighborhood of the Virgo Cluster. One of the goals for
future papers in this series will be the study of these two effects with
the denser set of 900 distance measurements now available.

Our program is complementary to the approach taken in the
IRAS/POTENT analysis (Bertschinger {\it et al.} 1990; Dekel {\it et al.}
1993), in which one predicts the mass distribution as a function of
position from peculiar velocity measurements and the Hubble and density
parameters. The results from these latter studies give preference for a
value of the density parameter in the range $0.6 < \Omega_0 < 1.3$. The
procedure starts from a map of the velocity field which is noisy because
of measurement uncertainties and requires substantial corrections for
biases and heavy smoothing. The POTENT corrections have been very
carefully tested in N-body models, however, and it would be exceedingly
useful to subject our approach to similar scrutiny. We have recently
begun this important exercise.

It is worth emphasizing that a value of the density parameter near
$\Omega =0.4$ would not strongly violate the constraints from either
approach.   Since the POTENT analysis is, of necessity, applied to a
larger region, the results are quite reconciled if there are
density fluctuations on scales $> 3000$ \kms\ that would not affect our
analysis.  Given the great differences in methodologies, the degree of
consistency of results from the two approaches is encouraging. Continued
comparison between the two methods should provide valuable insight into
the nature of the large-scale mass distribution and the total matter
content of the universe.

\subsection{Mass Bias and the Numerical Action Method}

If there is a significant mass component that is more smoothly
distributed than our mass tracers it will not be fairly measured by the
relative motions of nearby mass tracers, and the computation therefore
will underestimate the mean mass density. Branchini \&\ Carlberg (1994)
and Dunn \&\ Laflamme~(1995) give specific examples of this effect in
the biased CDM cosmogony. This cosmogony predicts that large galaxies
have overlapping halos. The result is very clearly illustrated in
Figure~5 of Branchini \&\ Carlberg (1994), which shows that for their
choice of parameters in the CDM cosmogony the effective mass derived
from the relative motions of galaxy pairs at separation $\sim 1$~Mpc is
only $\sim 20\%$ of the total within 5~Mpc of the pair, while at
separation $\sim 2$~Mpc the effective mass is close to the mass within
5~Mpc. These numbers depend on the shape and normalization of the mass
fluctuation power spectrum, but the effect surely is generic to the
picture. However, the effect is contrary to the evidence from the
redshifts of the galaxies in and near the Local Group. Zaritsky {\it et
al.} (1989) point out that if the radial velocities of the dwarf
spheroidal galaxy Leo~I at 200~kpc from the Milky Way and the Andromeda
Nebula at 700~kpc distance are the result of the gravitational assembly
of the Local Group, then the mass-to-light ratio of the two dominant
group members, the Milky Way and M31, is $M/L\sim 100$~solar units, and
the mass of the Milky Way is concentrated within a radius of $\sim
200$~kpc, contrary to the prediction of the CDM cosmogony. The same
result follows from the computation of orbits by the numerical action
method (Peebles 1994b). The action solutions also indicate that the
redshifts of the neighboring groups of galaxies, at distances $r\sim
3$~Mpc, require a similar value of the mass-to-light ratio (P94). That
is, the evidence is that most of the mass in the nearby groups is
concentrated in relatively compact halos.  There is no room for
substantially more mass in halos extending to 1~Mpc.

In any event, our mass tracers are groups of galaxies defined by a
procedure that minimizes the biases identified by Branchini \&
Carlberg and Dunn \& Laflamme.  By construction, none of our mass
tracers should have intersected: they are either moving apart or
approaching one another for the first time.  Hence, there should be
negligible overlapping halos, even if, contrary to the evidence
from the nearby region, galaxies have halos that extend beyond 1~Mpc
radius.

Perhaps some other form of dark matter has a broader coherence length,
so it is not detected in relative motions on scales of a few
megaparsescs. If so we may hope to detect it in more detailed sampling
of the peculiar motions within 3000~\kms. The present indications are in
the opposite direction, however, suggesting that $M/L$ is larger in the
most dense regions. That is, the mass distribution can be more strongly
clustered than the galaxies rather than smoother, at least in some
circumstances.  It is also remarkable that the value of the
mass-to-light estimated in  this much larger scale study is so similar
to the mass-to-light obtained in the above studies of indiviual groups
of galaxies.

Our use of the action principle assumes each galaxy or group of galaxies
in a mass tracer has behaved as a particle back to a time when the
galaxy peculiar velocities were much smaller than they are now. If two
galaxies merged at low redshift the assumption can still be a good
approximation when applied to the motion of the center of mass of the
material now in the single galaxy, if the two parts originated at
neighboring comoving positions. Our use of mass tracers that contain
several galaxies simply extends the merging to a larger scale. Although
we did not set out directly to test the effect of this merging we can
report the following experience. In early computations we used a lower
density threshold for the assignment of galaxies to mass tracers, so
that we had approximately half the number of mass tracers in the present
sample. This coarser sampling yielded very similar orbits and numerical
values for $\chi^2$ as a function of $M/L$ and $t_0$. Future
computations with faster machines and more efficient variational
techniques may allow another test based on finer sampling. If the mass
tracers are smaller, we will have to deal with a larger number of
triple-value and multiple-value regions.  This investigation can be
done, perhaps following the method in Peebles (1994b), but it will be a
demanding computation. Perhaps a better short-term hope for a more
complete test is the use of realistic N-body simulations, as in the
analysis by Branchini and Carlberg (1994) of a nearly unbiased numerical
simulation. It will be possible to test for the effects of non-spherical
mass tracers, possibly an important consideration at early times, and
for the influence on the determination of $\Omega_0$ of what Dunn \&
Laflamme (1995) call the `orphan' particles of a CDM-type model, the
most weakly clustered component.

\subsection{Large-Scale Mass Fluctuations}

Our generalization of the numerical action method to take account of
large-scale ($> 3000$ \kms) density fluctuations by an imposed
gravitational field that evolves in accordance with perturbation theory
is meant to probe the nature of the large-scale density fluctuations and
to remedy the aspherical discontinuity resulting from the redshift cut
in the sample of mass tracers. In the numerical results shown in \S 5,
the application of the gravitational field of an external mass
distribution modeled as the distribution of the Abell clusters actually
improves the consistency of the computed orbits with the observations.
There is a delicate balance, however, because with such large mass
assigned to each cluster a small change in position of one of the nearer
ones can have a large influence on the orbits of the mass tracers. A
better model would put some fraction of the mass in the great clusters
and the rest at, say, the positions of IRAS galaxies. We will
investigate such a model in the next phase of our study.  It will be
interesting to see whether a believable measurement of this mass
fractionation between great clusters and spirals can be derived from the
behavior of the $\chi^2$ statistic.

\section{Acknowledgments}

We thank the following people for making digital versions of their
catalogs available to us: Marc Davis and Michael Strauss for the 1.9 Jy
IRAS Survey, Nick Kaiser for the binned 1-in-6 0.6 Jy IRAS survey, Tod
Lauer and Marc Postman for their Abell cluster sample and, especially,
Amos Yahil and Karl Fisher for a pre-release version of the 1.2 Jy IRAS
survey, from which we extracted sources with $cz < 3000$ \kms.
We thank Adrian Melott for discussions. This
research was supported in part by grants from the National
Science Foundation and at Caltech by the Fairchild Distinguished Scholar
Program.

\eject

\clearpage

\begin{figure}

\caption {The distribution of blue light associated with galaxies in
sin~$|b|$ bins (where $b$ is galactic latitude; bin increments are 0.02
in sin~$|b|$). The solid histogram represents the 3030 galaxies in the
merged NBG catalog and IRAS sample. The dashed histogram illustrates the
situation with the inclusion of 151 `fake' entries at $|b| <
5^{\circ}$. Appropriately, there is rough constancy of luminosity in
sin~$|b|$ bins in the combined data set of real plus fake mass tracers.
} \label{catdis}

\caption {Histogram of the number of mass tracers per luminosity bin.
Solid histogram is based on the entire catalog.  There is a sharp
turnover because of incompleteness at the faint end and because the
objects at large $cz$ are boosted in luminosity by the correction
factor.  Orbits are calculated for mass tracers with $L_B^{b,i} > 5
\times 10^9 L_{\odot}$ and/or with measured distances. The dotted
histogram is restricted to $cz < 1500$ \kms\ and $|b| >
30^{\circ}$, where the catalog is more nearly complete and the
corrections for luminosity are small.  The turnover at $10^9 L_{\odot}$
is probably due to residual incompleteness. } \label{lumbin}

\caption {The histogram of the Lauer-Postman sample of Abell clusters in
sin~$|b|$ bins (bin increments are 0.05 in sin~$|b|$). `Fake' sources
are generated from information provided in the low latitude distribution
of the QDOT 0.6 Jy IRAS survey but cannot sensibly be plotted in this
figure because they are all located in the single bin associated with $b
= \pm 10^{\circ}$. } \label{lpdis}

\caption {The boundary of a velocity-limited survey in a plane
containing the peculiar motion of the observer.  Different line types
refer to ratios of peculiar velocity of the observer to the cutoff
redshift of (0.0, 0.2, 0.4, 0.6, 0.8, and 1.0).  In the case of the
Local Group this ratio is $\sim 0.1$, hence the distortion is minor
though the origin is shifted. } \label{sampleshape}

\caption {Action orbits.  The model used for this illustration has $M/L
= 200$, $t_0 = 9.7$~Gyr, and excludes sources beyond 3000 \kms.  The
coordinates are supergalactic and only tracers within $\pm 5$ Mpc of the
equatorial plane are plotted.  Diamonds locate model positions today.
Curved tails show the orbits of tracers drawn from the Coma-Sculptor
Cloud (NBG catalog terminology).  There are fat tails in the 3 cases of
the Local Group (at origin), the Virgo Cluster (upper left quadrant),
and the Ursa Major Cluster (upper right quadrant). } \label{tadpoles}

\end{figure}

\begin{figure}

\caption {Contours of $\chi^2$ as a function of the two free parameters
$M/L$ and $t_0$.  This reduced $\chi^2$ is calculated from the
difference between model and observed distance moduli for 289 galaxies
with luminosity-linewidth distance measurements.  The normalization is
given by the expected measurement error of $0.4^m$. Contours are at
intervals of $0.08 \simeq 1\sigma$. The dashed line goes along the
bottom of the minimum $\chi^2$ valley.  The lines of constant $H_0$ and
$\Omega_0$ are based on the Loveday {\it et al.} luminosity density
normalization. ($a$) No external perturbation by masses beyond 3000
\kms\ distance. ($b$) External perturbation modeled by the distribution
of the great clusters. } \label{chi2map}

\caption {Minimum reduced $\chi^2$ as a function of time (dashed line,
read from right axis), and mass-to-light ratio at the minimum $\chi^2$
(solid line, read from left axis).  These values track the floor of the
$\chi^2$ valley in the previous figure. ($a$) No external sources
considered. ($b$) External sources taken into account.  In this latter
case, the $\chi^2$ minimum is deeper and better defined. }
\label{minchi}

\caption{Observed minus model distance moduli versus redshift for a
model without external sources.  In this case, $M/L = 150$ and $t_0 =
10.3$. There tends to be a systematic decrease toward higher $cz$. If
$M/L$ choices are too high or too low the points systematically shift
down or up, respectively. } \label{delmu_cz_noex}

\caption{Observed minus model distance moduli versus redshift including
external sources. Now, systematics are significantly reduced.  $M/L =
160$, $t_0 = 9.8$, and $\chi^2 = 1.20$. }\label{delmu_cz}

\caption{Observed minus model distance moduli versus redshift including
external sources but at a higher mass-to-light choice along the minimum
$\chi^2$ valley. $M/L = 500$, $t_0 = 8.0$, and $\chi^2 = 1.54$.   The
rms scatter is consistently larger than that in Figure~9.
}\label{delmu_cz2}

\caption{Observed peculiar velocities ($U_{obs} = cz - H_0 D_{TF}$)
versus model peculiar velocities ($U_m = cz - H_0 D_{m}$) for the case
with $\Omega_0 = 0.10$ and $H_0 = 90$ and with external masses included.
A linear regression fit (dashed line) is made with errors in the
observed quantities.  The oblique solid line is where $U_{obs} = U_m$.
Errors are derived from an assumed rms uncertainty of $0.4^m$ in an
individual distance modulus.  The axes have different scales to
accommodate the much greater scatter in the observed quantities. }
\label{um_u0}

\end{figure}

\begin{figure}

\caption{Observed minus model peculiar velocities versus model peculiar
velocities.  This plot provides another way of viewing the data of the
previous figure.  The regression fit (dashed line) is close to
horizontal as it should be for a good model. } \label{dUvsU}

\caption{Cumulative luminosity density profiles as a function of
distance from the Local Group. The horizontal lines with the same type
style show the mean luminosity density (which differs between models,
depending on $H_0$). The solid curve corresponds to a case near the
minimum of $\chi^2$: $M/L = 200$ and $t_0 = 9.7$ Gyr. Here the
luminosity density just approaches the global mean at the edge of the
sample region. The situation is similar with other cases near the
$\chi^2$ minimum valley (dots: $M/L=45$ and $t_0 = 11.4$ Gyr; short
dashes: $M/L=475$ and $t_0 = 8.4$ Gyr).  In models off the $\chi^2$
minimum valley the luminosity density can fall more quickly (long
dashes: $M/L=800$ and $t_0= 9.7$ Gyr) or more slowly (dash-triple dots:
$M/L=50$ and $t_0= 9.7$ Gyr). }\label{density}

\caption{Two projections of the peculiar velocity of the Local Group in
supergalactic coordinates.  Solid lines connect results from models
along the $\chi^2$ minimum valley in the case of no external masses;
dashed lines connect results with external masses. Each ``+'' represents
the result from a specific model, identified by the value of the
expansion time $t_0$. The large plus signs show the motion of the Local
Group with respect to the microwave background radiation. }
\label{lgvel}

\end{figure}
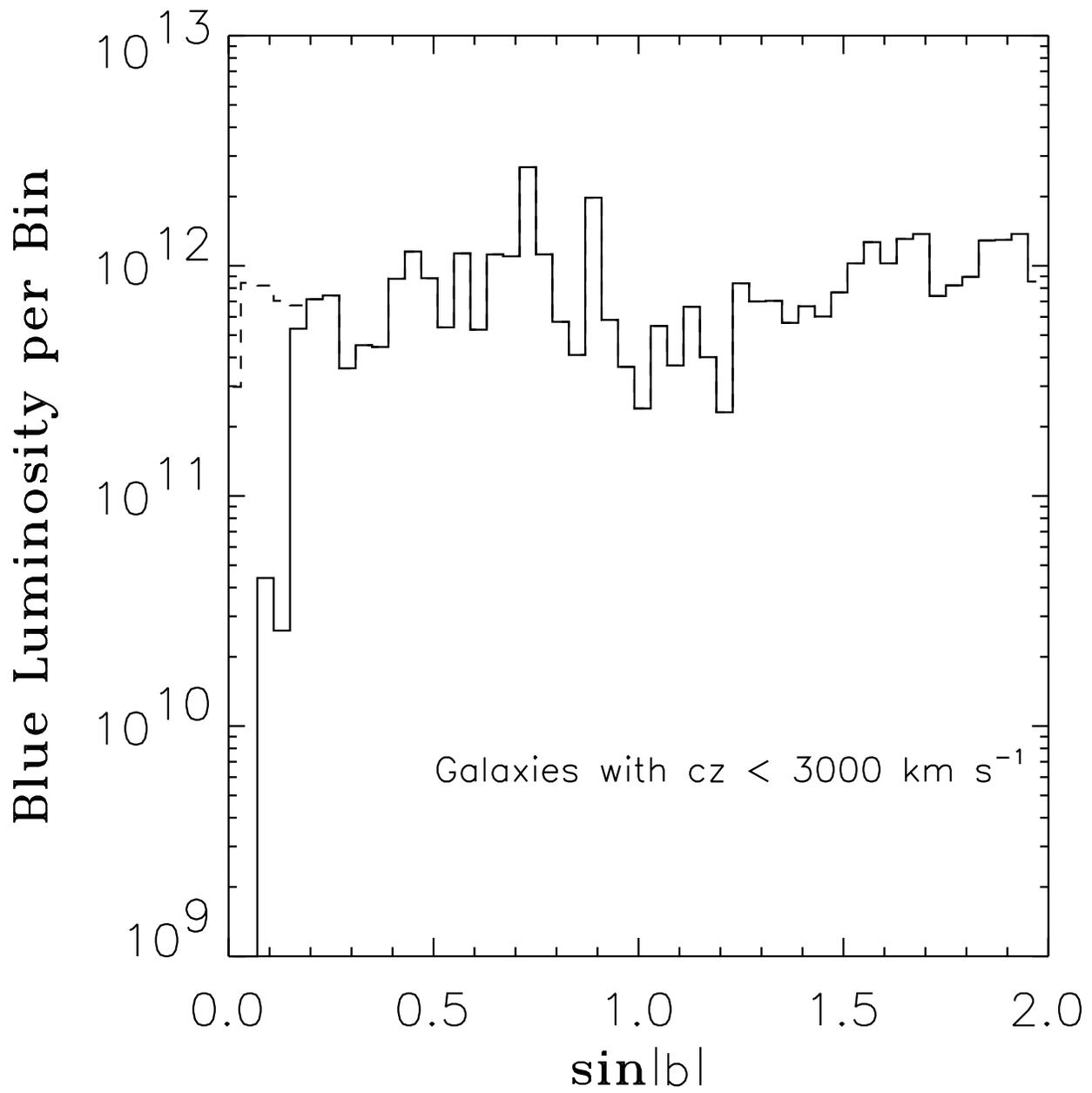
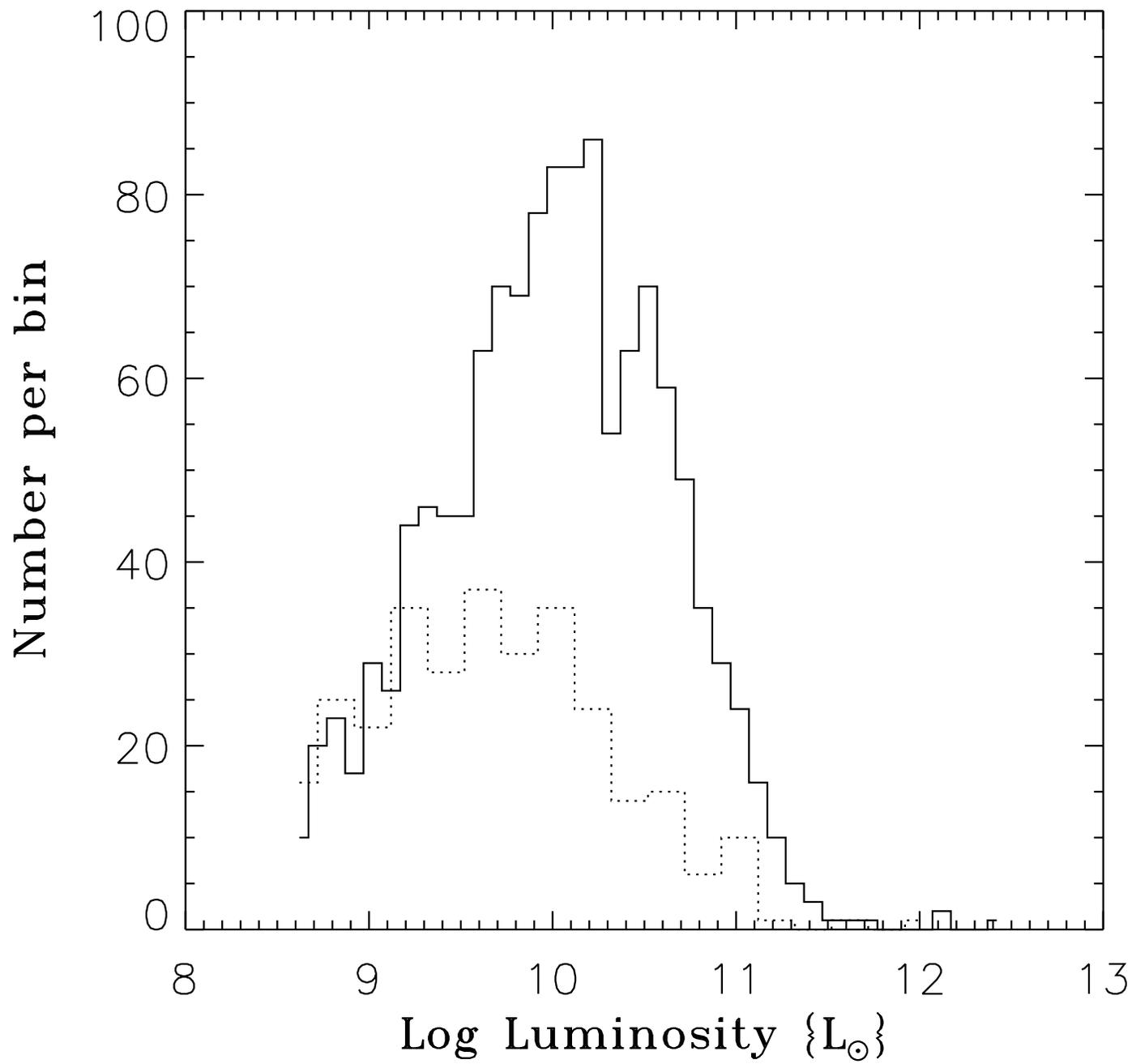
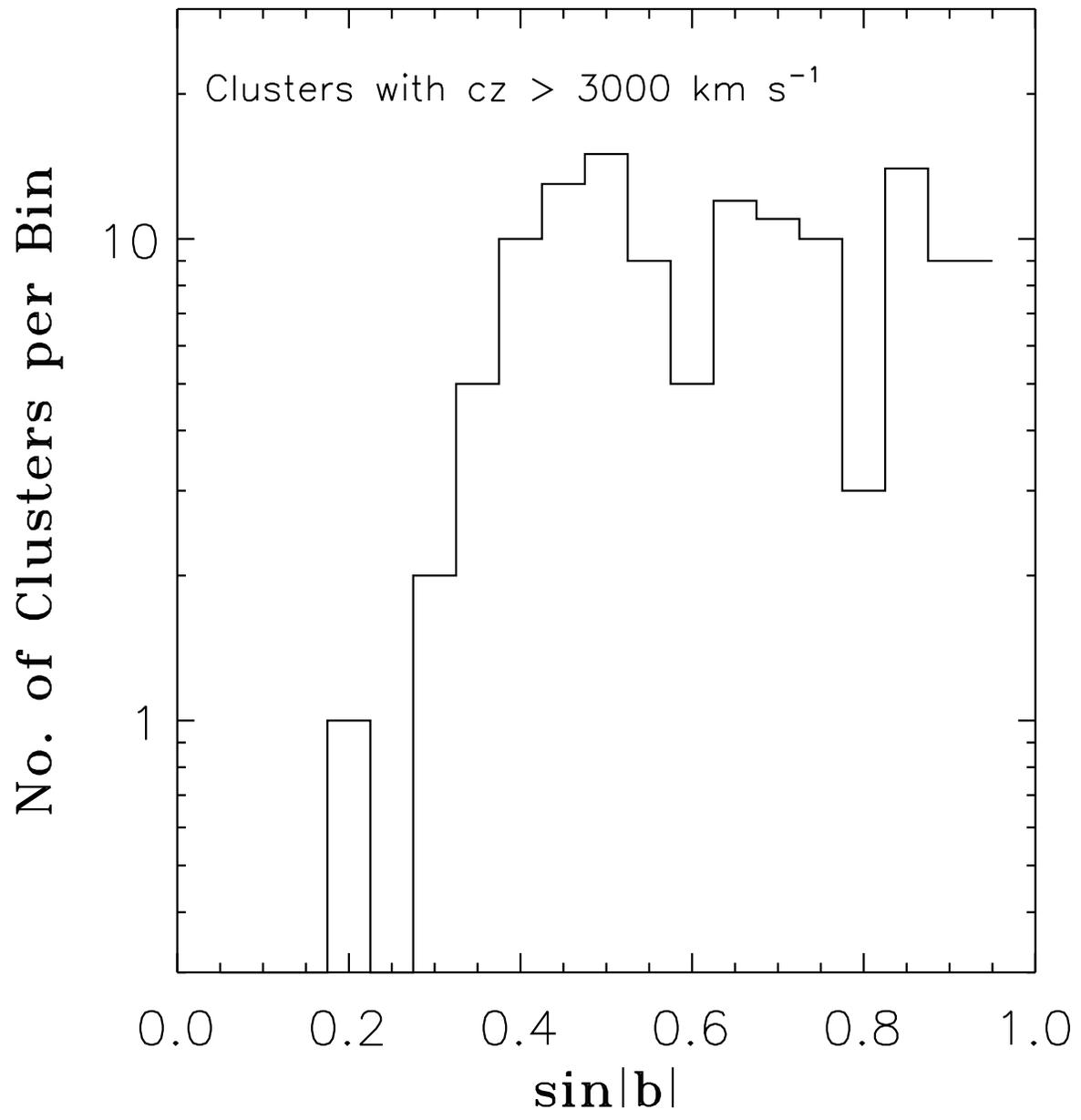
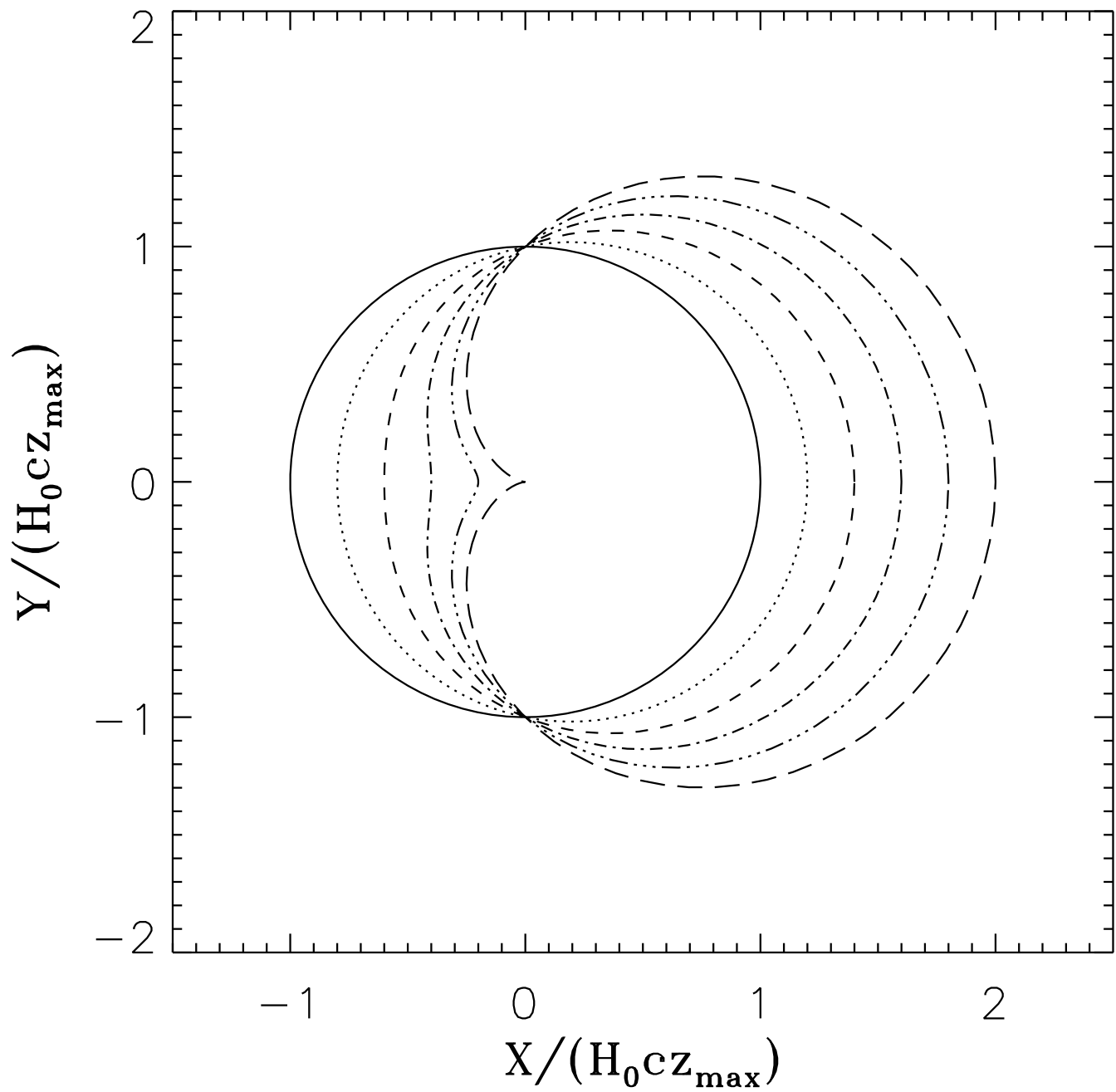
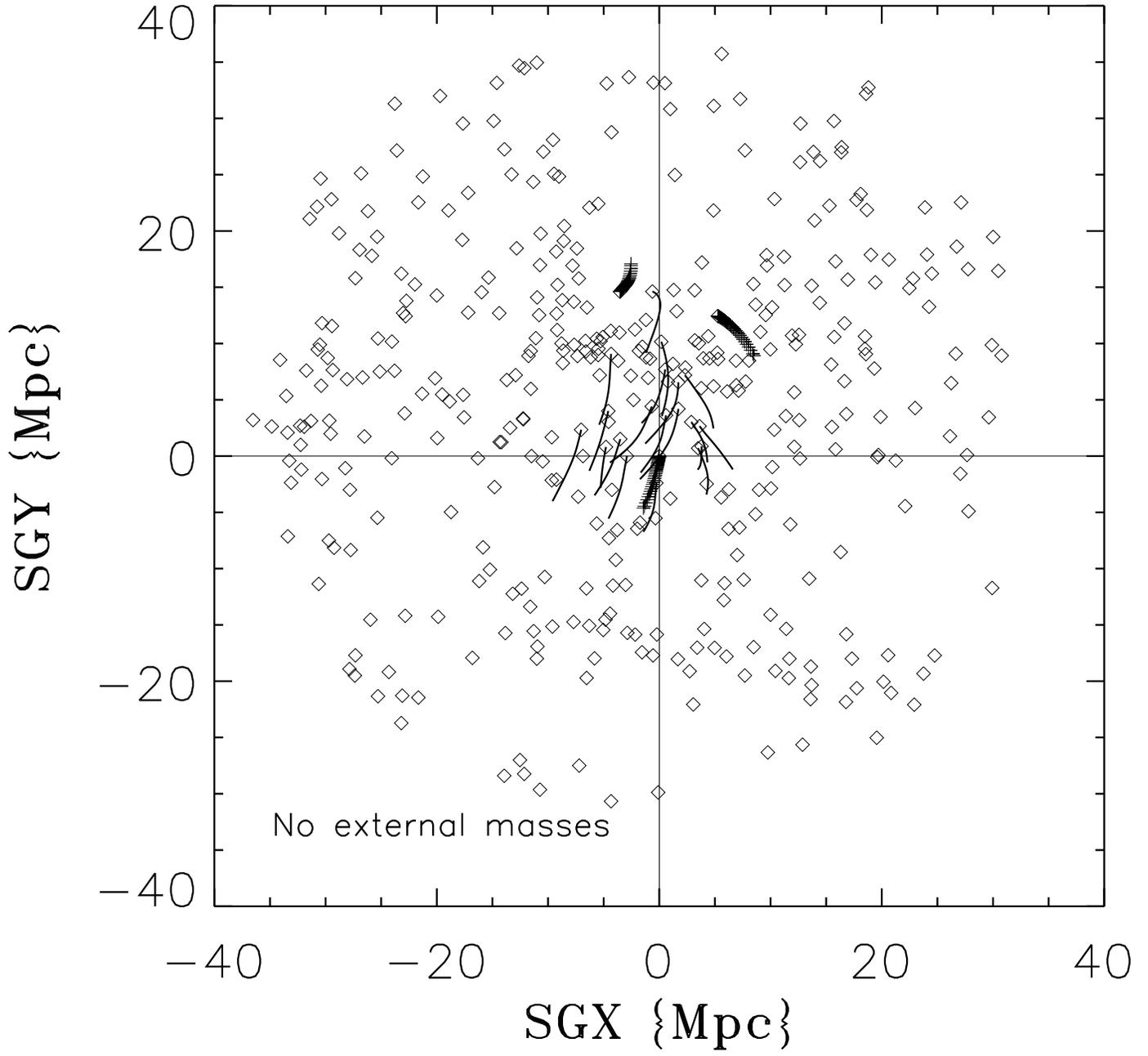
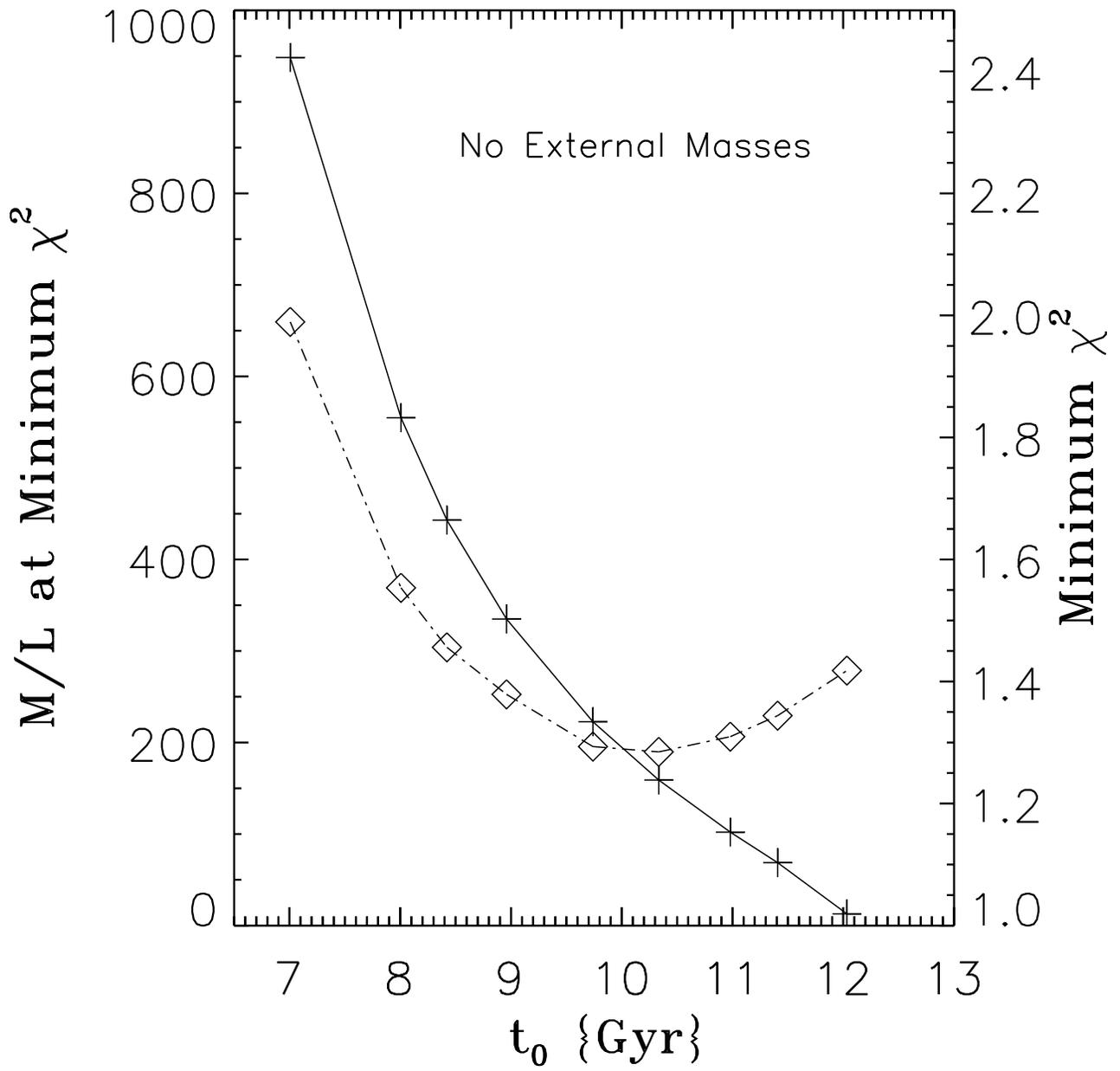
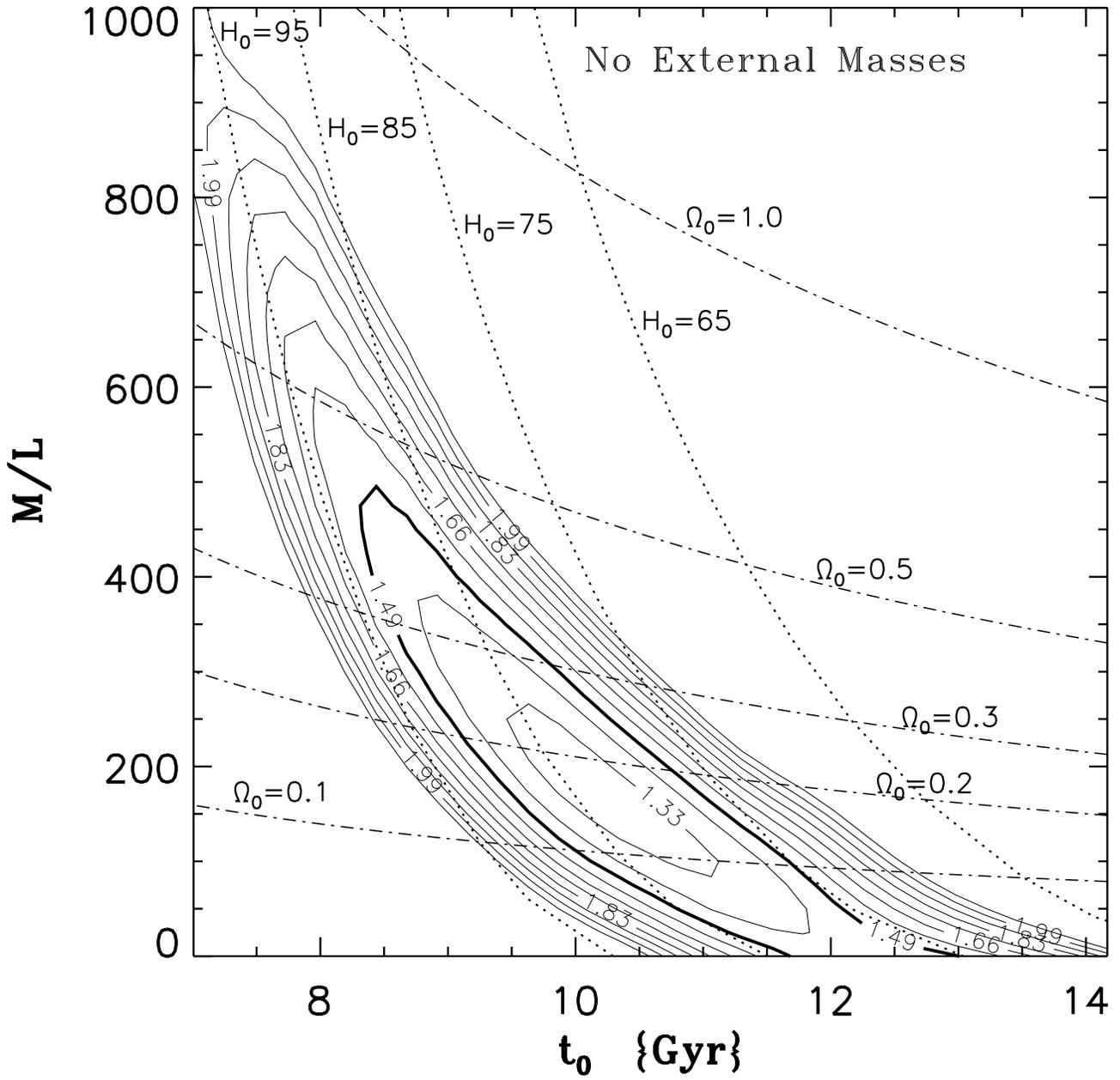
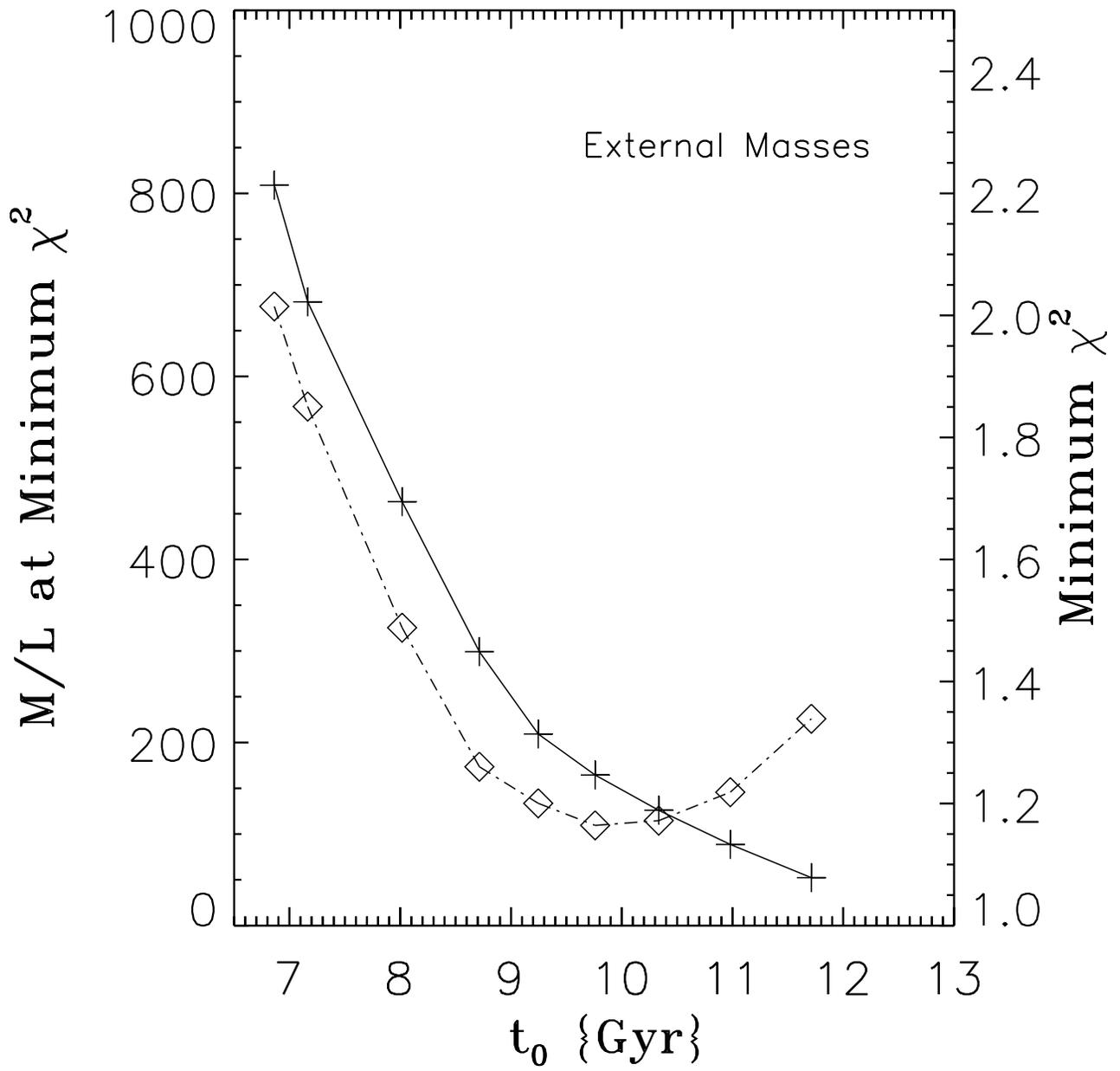
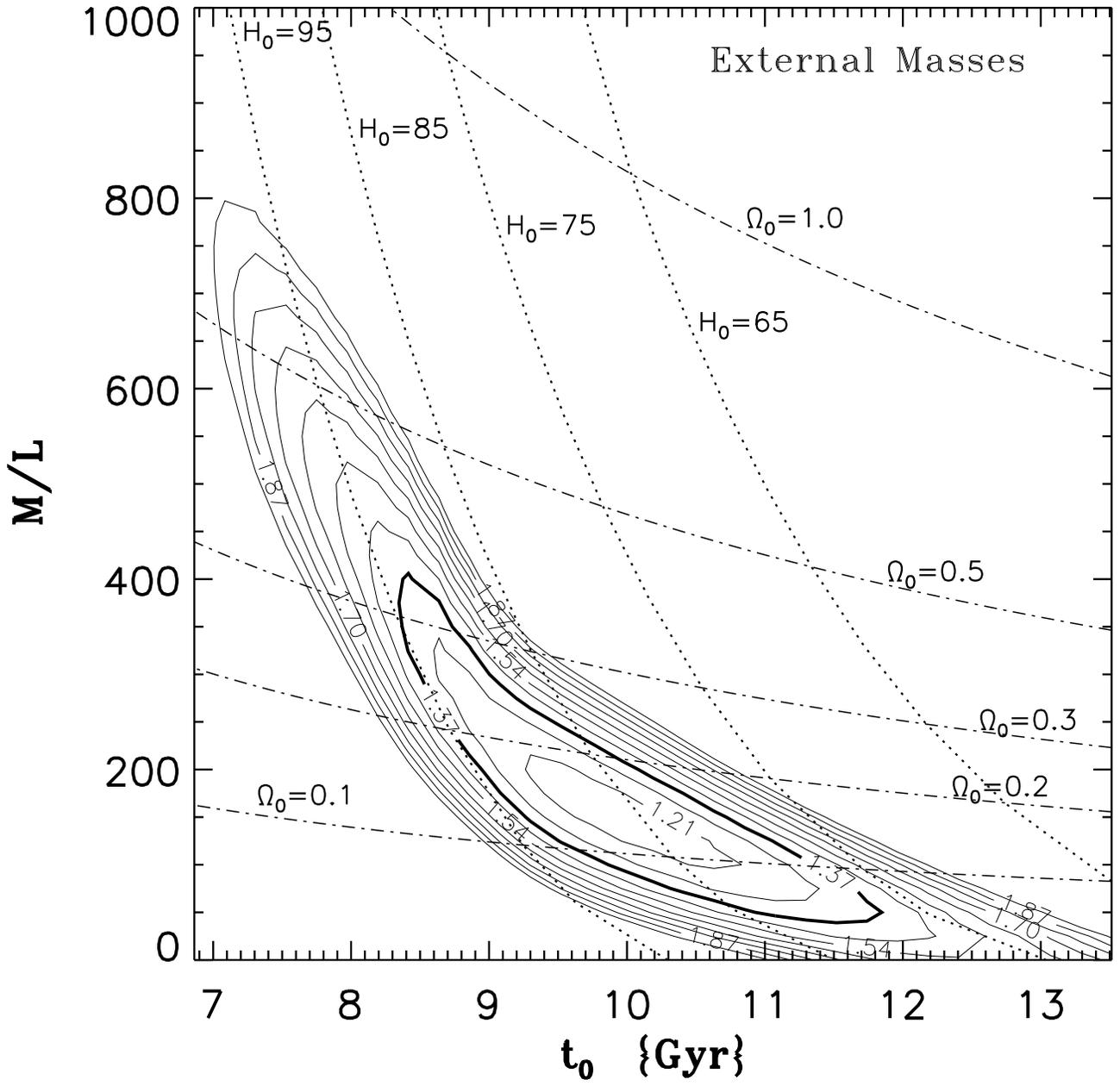
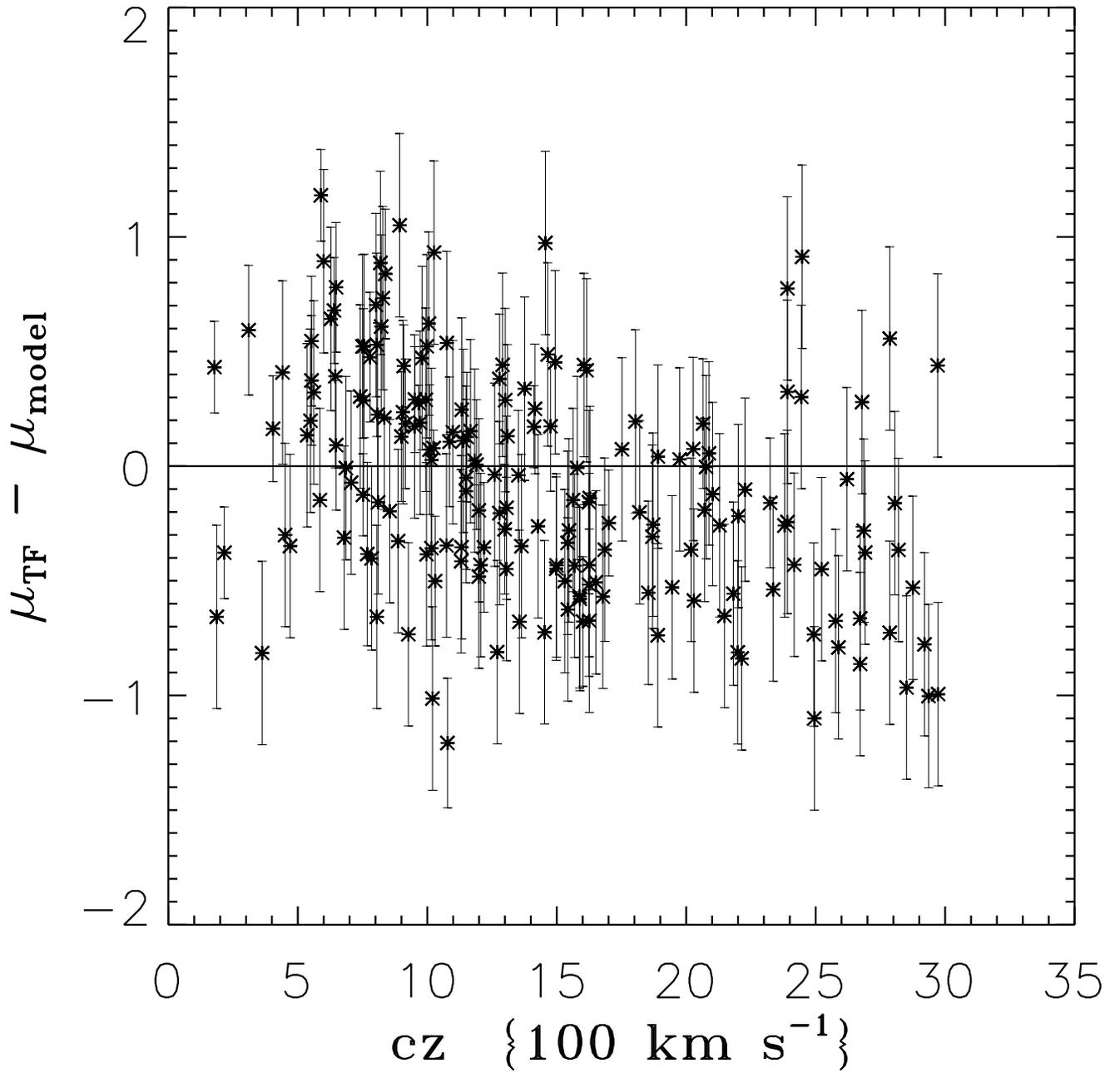
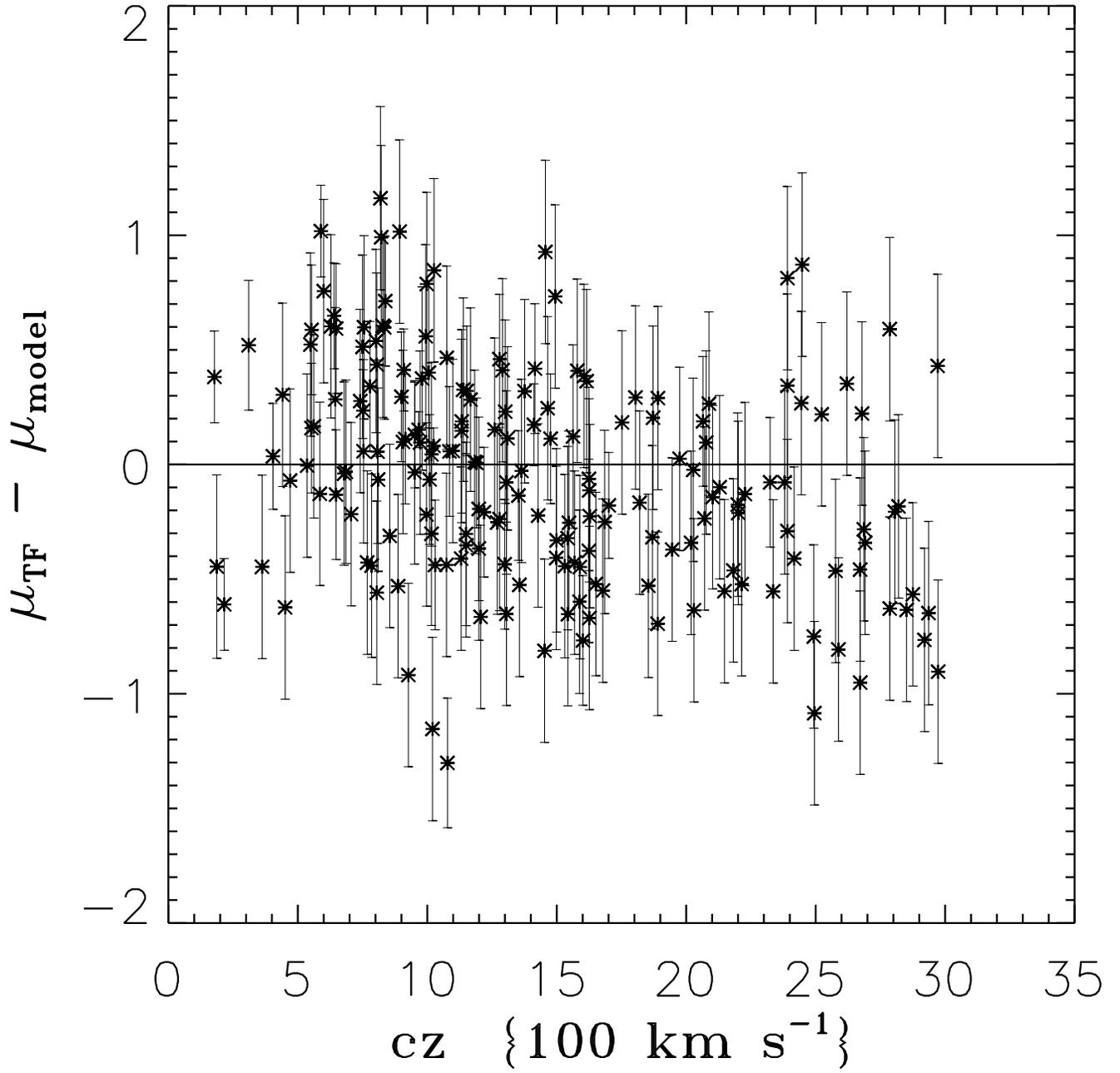
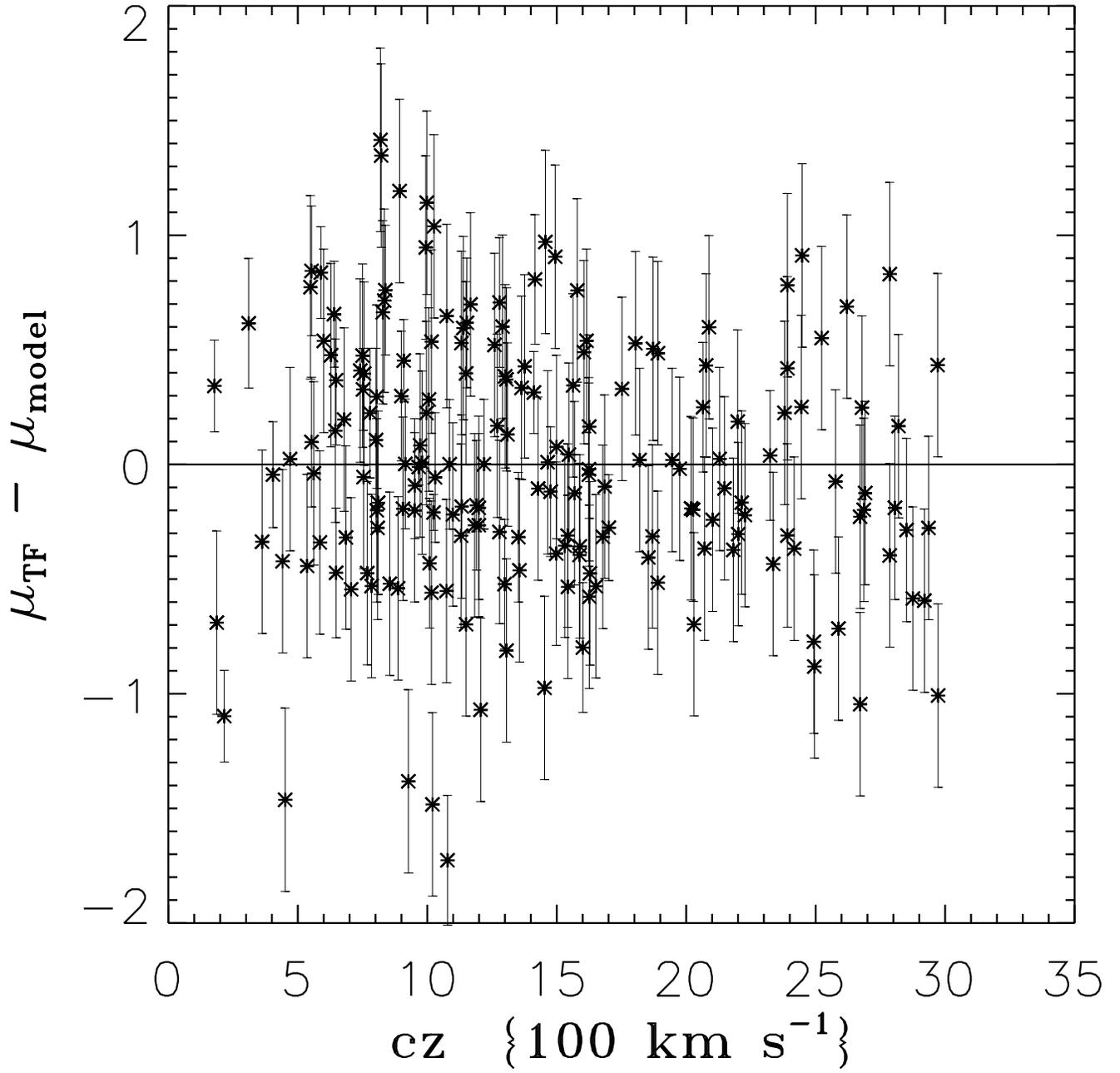
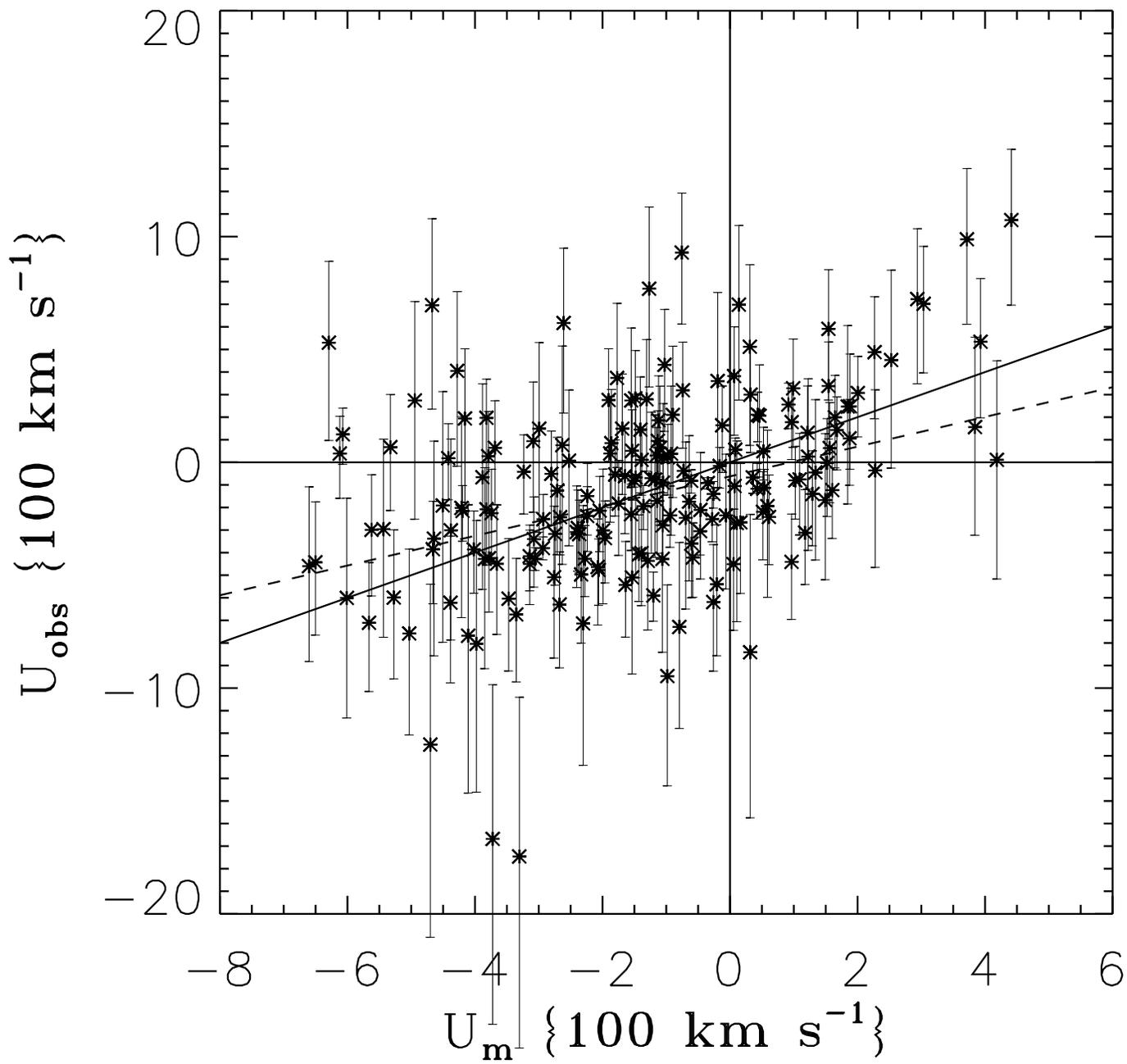
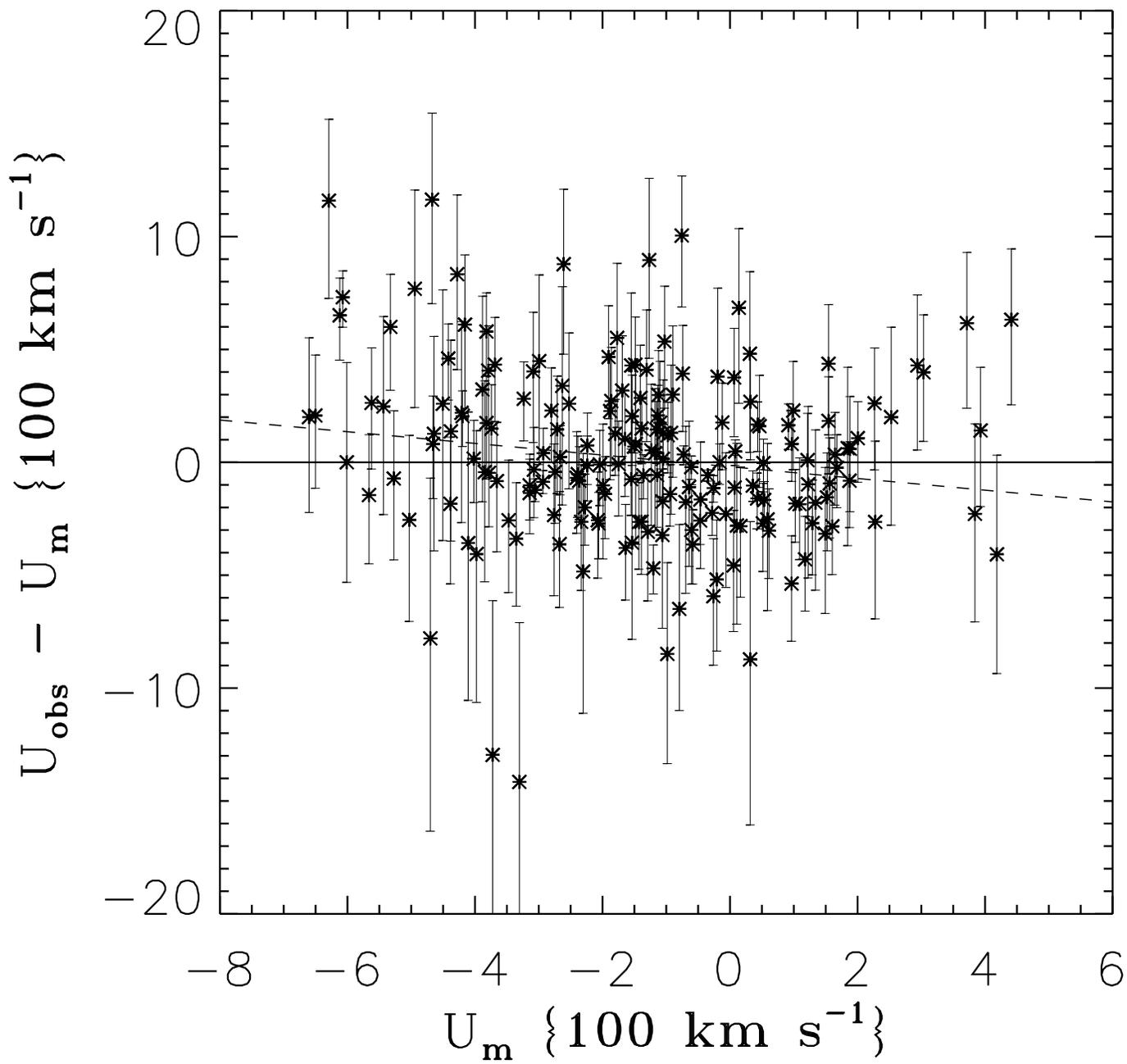
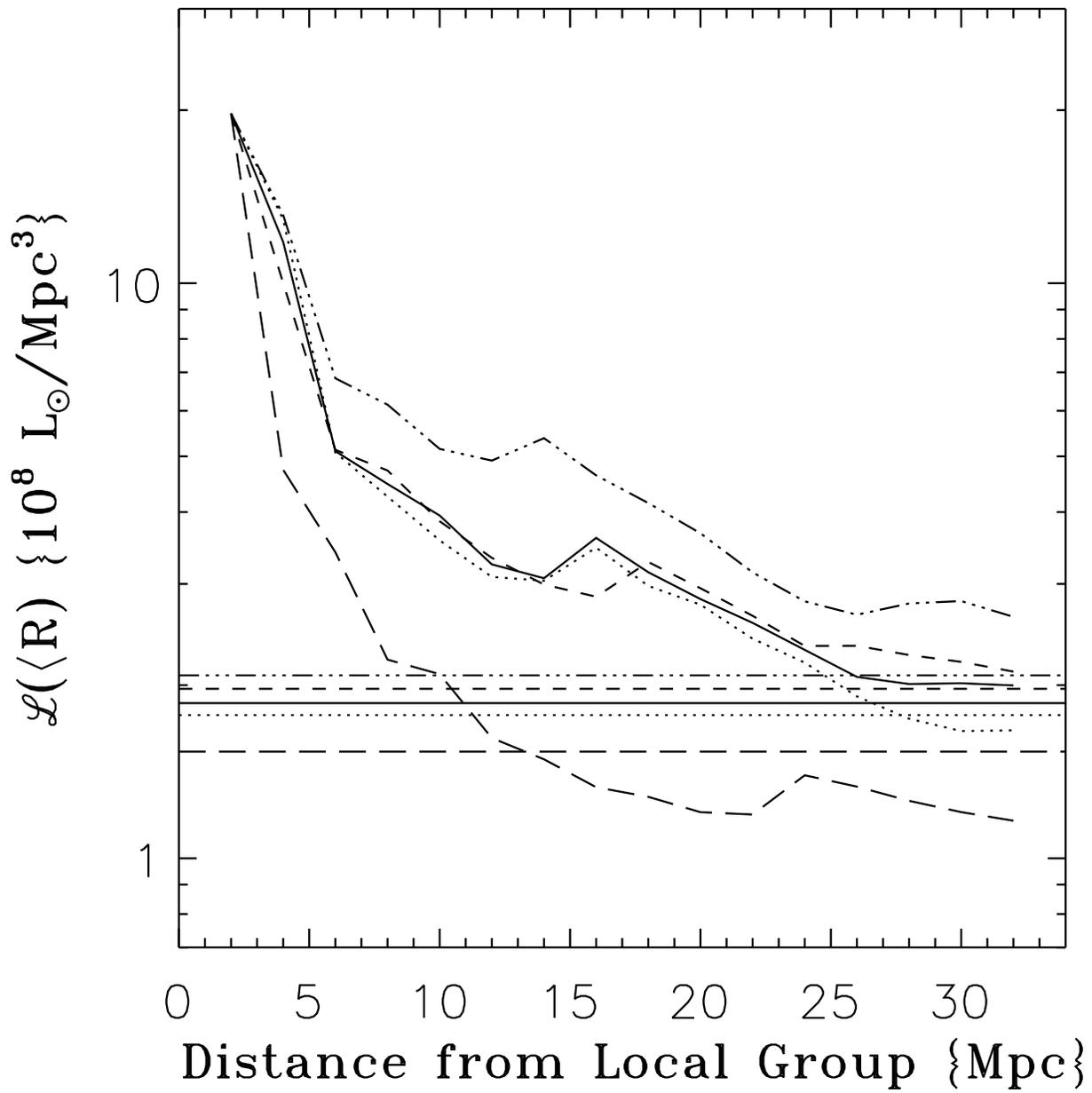
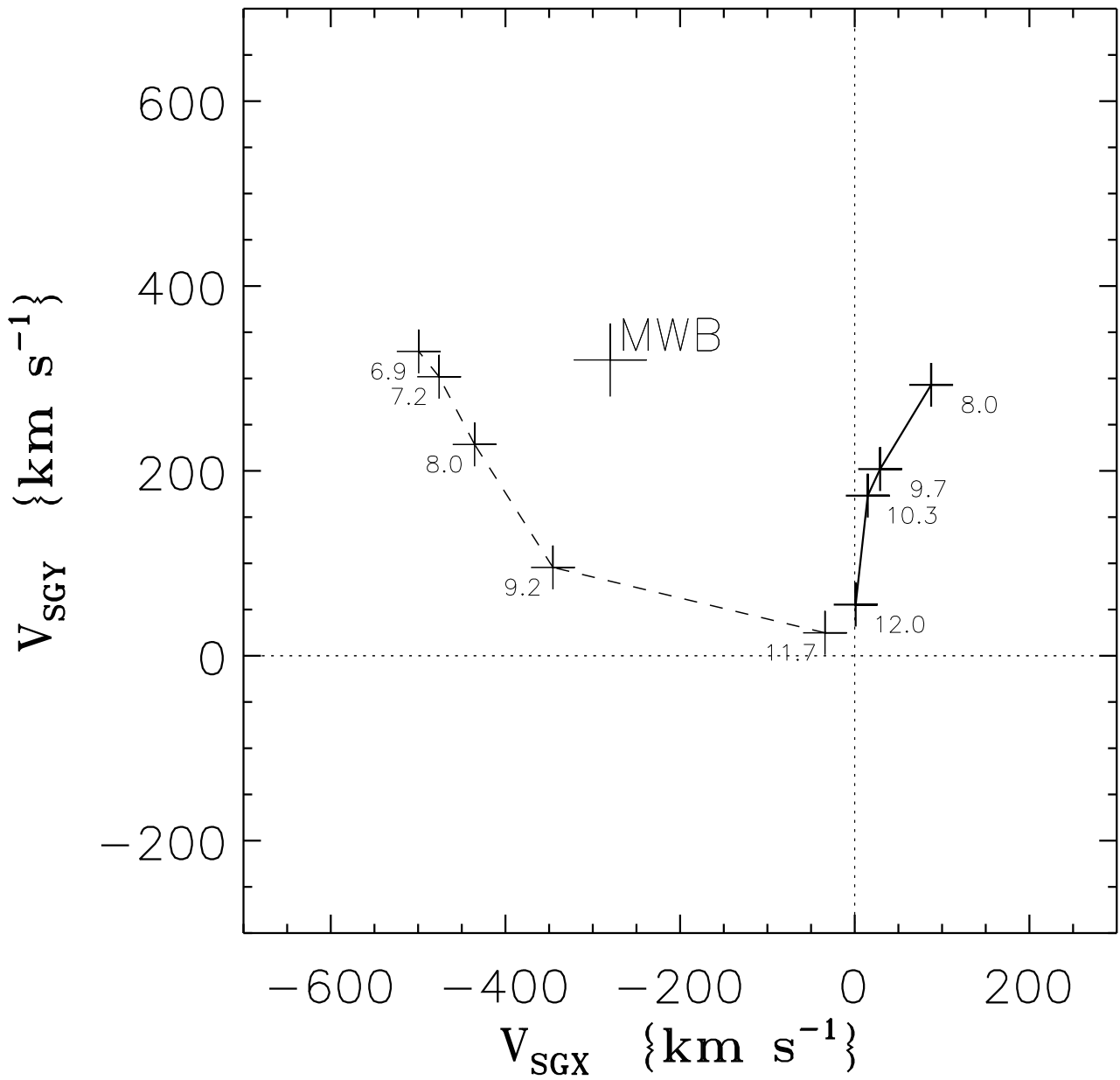
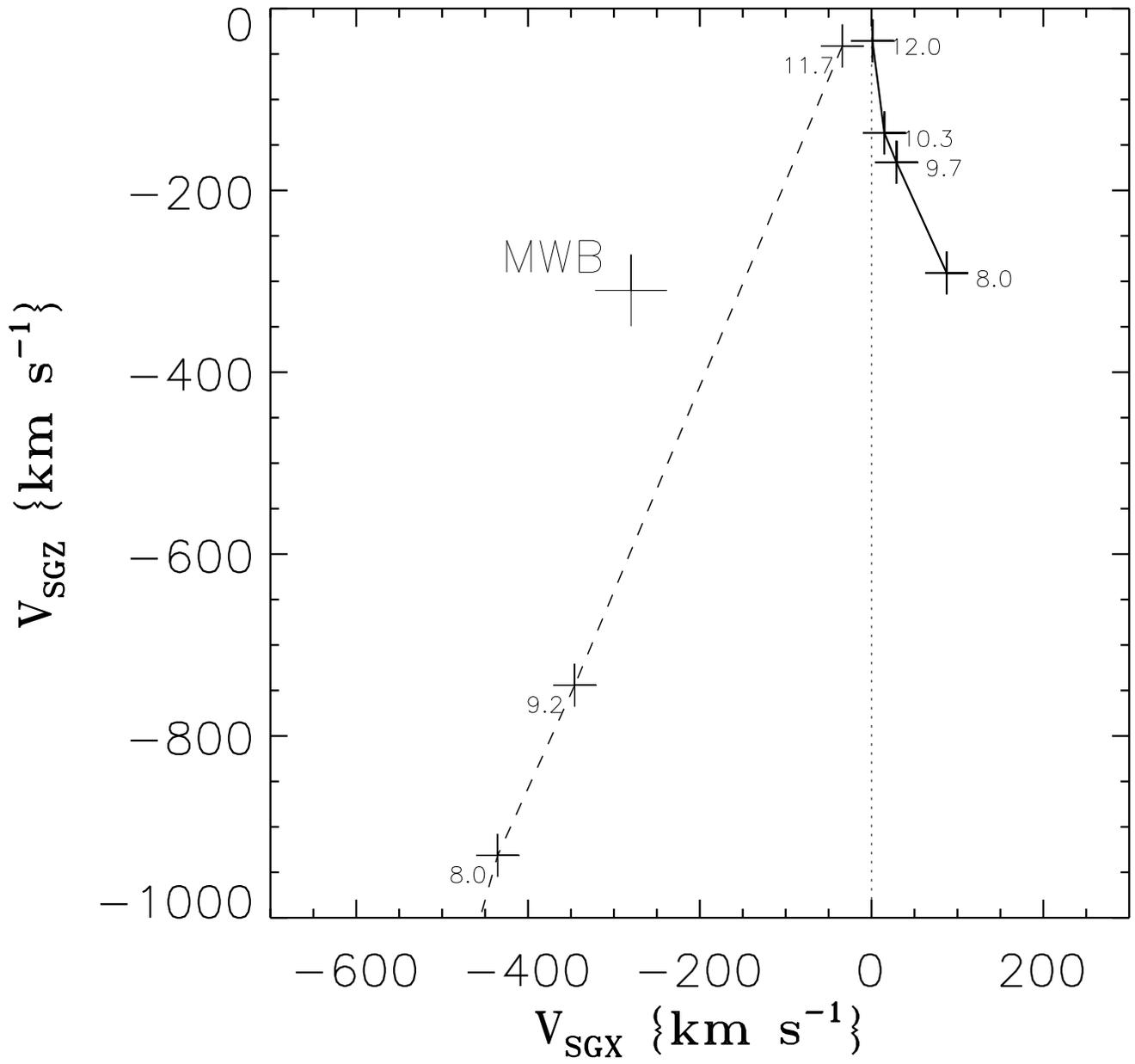
\end{document}